\documentclass[journal]{IEEEtran}

\usepackage{cite}

\usepackage{makeidx,epsfig}
\usepackage{setspace,graphicx,multirow}

\usepackage{amsfonts,latexsym,amssymb,amsthm}

\usepackage{graphicx}
\usepackage{epstopdf}

\usepackage{caption3} 
\DeclareCaptionOption{parskip}[]{} 
\usepackage[small]{caption}

\usepackage{subcaption}

\usepackage{array}

\usepackage[cmex10]{amsmath}
\DeclareMathOperator{\diag}{diag}  
\DeclareMathOperator{\var}{var}

\usepackage{url}

\usepackage{multirow}
\usepackage{hhline}

\usepackage{amsmath}
\usepackage{algorithm}
\usepackage[]{algpseudocode}
\usepackage{varwidth}

\makeatletter
\def\BState{\State\hskip-\ALG@thistlm}
\makeatother

\usepackage{algcompatible}

\usepackage{fixltx2e}

\newcommand*\diff{\mathop{}\!\mathrm{d}}

\usepackage{enumitem}

\makeatletter
  \newcommand\tinyv{\@setfontsize\tinyv{7pt}{9}}
\makeatother

\usepackage{authblk}

\usepackage{textcomp}

\usepackage{dsfont}

\usepackage{color}

\ifodd 0
\newcommand{\rev}[1]{{\color{red}#1}} 
\newcommand{\com}[1]{\textbf{\color{blue} (COMMENT: #1)}} 
\else
\newcommand{\rev}[1]{#1}

\newcommand{\com}[1]{}
\fi

\begin{document}
\bibliographystyle{IEEEtran}
\bstctlcite{IEEEexample:BSTcontrol}

\title{Hybrid Active/Passive Wireless Network Aided by Intelligent Reflecting Surface: System Modeling and Performance Analysis}

\author{Jiangbin~Lyu,~\textit{Member,~IEEE},
        and~Rui~Zhang,~\textit{Fellow,~IEEE}%
\thanks{This work was supported in part by the National Natural Science Foundation 
	of China (No. 61801408 and No. 61771017), the Natural Science Foundation of Fujian 
	Province (No. 2019J05002), the Fundamental Research Funds for the Central 
	Universities (No. 20720190008), and the National University of Singapore under Research Grant R-261-518-005-720 and R-263-000-E86-112.} 
\thanks{J. Lyu is with the School of Informatics, and Key Laboratory of Underwater Acoustic Communication and Marine Information Technology (Ministry of Education), Xiamen University, China 361005 (e-mail: ljb@xmu.edu.cn). R. Zhang is with the Department of Electrical and Computer Engineering, National University of Singapore, Singapore 117583 (email: elezhang@nus.edu.sg).}
\thanks{Copyright \textcopyright 2021 IEEE. Personal use of this material is permitted. However, permission to use this material for any other purposes must be obtained from the IEEE by sending a request to pubs-permissions@ieee.org.}%
}


\markboth{IEEE Transactions on Wireless Communications}%
{To appear}
%



\maketitle

\begin{abstract}

Intelligent reflecting surface (IRS) is a new and promising paradigm to substantially improve the spectral and energy efficiency of wireless networks, by constructing favorable communication channels via tuning massive low-cost passive reflecting elements.
Despite recent advances in the link-level performance optimization for various IRS-aided wireless systems, it still remains an open problem whether the large-scale deployment of IRSs in wireless networks can be a cost-effective solution to achieve their sustainable capacity growth in the future.
To address this problem, we study in this paper a new hybrid wireless network comprising both active base stations (BSs) and passive IRSs, and characterize its achievable spatial throughput in the downlink as well as other pertinent key performance metrics averaged over both channel fading and random locations of the deployed BSs/IRSs therein based on \textit{stochastic geometry}.
Compared to prior works on characterizing the performance of wireless networks with active BSs only, our analysis needs to derive the power distributions of both the signal and interference reflected by distributed IRSs in the network under spatially correlated channels, which exhibit channel hardening effects when the number of IRS elements becomes large.
Extensive numerical results are presented to validate our analysis and demonstrate the effectiveness of deploying distributed IRSs in enhancing the hybrid network throughput against the conventional network without IRS, which \textit{significantly boosts the signal power} but results in only \textit{marginally increased interference} in the network.
Moreover, it is unveiled that there exists an \textit{optimal IRS/BS density ratio} that maximizes the hybrid network throughput subject to a total deployment cost given their individual costs, while the conventional network without IRS (i.e., zero IRS/BS density ratio) is generally suboptimal in terms of throughput per unit cost.
\end{abstract}

\section{Introduction}

%
%
%
%
%
%
%

The proliferation of mobile applications and explosive growth of wireless data have been continually spurring enthusiasm in inventing new and innovative wireless communication technologies that would achieve higher spectral/energy efficiency (SE/EE) yet at an affordable deployment/operational cost.
Among others, some prominent wireless technologies proposed in the last decade include ultra-dense network (UDN), massive multiple-input multiple-output (MIMO), and millimeter wave (mmWave) communication\cite{SmallCell5G}.
Although these technologies significantly enhanced the wireless network SE, they also incurred increasingly more energy consumption and higher hardware cost, due to the deployment of more base stations (BSs) and/or relays in the network as well as mounting them with more active antennas requiring costly radio frequency (RF) chains, especially when operating at mmWave frequency bands.
Therefore, it is doubtful whether the existing paradigm by adding more and more active nodes/components in the wireless network would be a cost-effective solution to achieve its sustainable capacity growth in the future.

To tackle this challenge, intelligent reflecting surface (IRS) has recently emerged as a promising solution based on the new concept of reconfiguring the wireless propagation environment that is traditionally deemed to be random and uncontrollable\cite{QQirsMag,IRSbasar,IRSmarcoCome,IRSholographic}.
Specifically, IRS is a planar surface consisting of a massive number of low-cost passive reflecting elements
that can be tuned dynamically to alter the amplitude and/or phase of the signal reflected by them, thus collaboratively reconfiguring the signal propagation to achieve various desired functions such as three-dimensional (3D) passive beamforming, spatial interference nulling and/or cancellation.
Compared to the conventional active relaying/beamforming, IRS does not require any active RF chain for signal transmission/reception but simply leverages passive wave reflection, thus leading to much lower hardware cost and energy consumption yet operating spectral efficiently in full-duplex (FD)
without the need of costly self-interference cancellation (SIC)\cite{QQirsMag}.
Moreover, IRS can be easily attached to or removed from the existing objects in the environment (e.g., walls and ceilings), and seamlessly integrated into cellular or WiFi systems without the need to modify their current infrastructure and operating standards\cite{QQirsMag}.
As such, IRS can be densely deployed in wireless networks at a low and scalable cost as well as with high flexibility and compatibility.

The appealing advantages of IRS have attracted a great deal of interest recently in investigating IRS-aided wireless systems from various aspects and/or under different setups, such as passive
beamforming design \cite{QQtwc,IRSjinShi,IRSschoberICCC,IRSqqDiscrete,IRSshuowen,IRScuiYing,IRSchauYuen}, IRS-aided orthogonal frequency division multiplexing (OFDM)
system \cite{IRSyifeiOFDM}\cite{IRSzhengBeixiong}, non-orthogonal multiple access (NOMA)\cite{IRSyangGangNOMA,IRSdingZhiguoNOMA,IRSbeixiongNOMA}, physical layer security \cite{IRSsecureGuangchi,IRSsecureYCliang,IRSsecureSchober,IRSsecureXinrong}, wireless information and power transfer \cite{QQirsSWIPTwcl,IRSnalanSWIPT,QQirsSWIPTqos}, and so on.
The above works on IRS-aided wireless systems mainly aim to optimize the system performance at the link level with one or more IRSs at fixed locations, which show that the IRS-aided system can achieve significant energy efficiency\cite{IRSchauYuen} and/or spectral efficiency improvement over the traditional system without IRS, with optimized IRS reflection coefficients.
In \cite{HanzoMIMOstochasticGeometry}, the authors investigate a multi-user system aided by multiple intelligent surfaces (equivalent to a single large IRS) co-located at a random location in the network.
However, since IRS typically serves users in its proximity, distributed IRSs should be deployed in the network to serve distant groups of users and thereby boost the network throughput.
Motivated by this, in our prior work\cite{IRSlyuSingleCell}, the spatial throughput of a single-cell multi-user system aided by distributed IRSs located at random locations is characterized, which is compared favorably with the conventional system aided by distributed relays but with significantly reduced active antennas, under their respectively optimized deployment.

Furthermore, for large-scale deployment of IRSs in future wireless systems, one critical issue is the modeling, design and performance characterization of the IRS-aided multi-cell hybrid wireless network comprising both distributed active BSs and passive IRSs subjected to the inter-cell interference.
There have been some recent works (e.g., \cite{IRSmulticellNallan,IRSmulticellXujie,IRSmulticellCuiYing,IRSdistributedMultiPairsShiYuanming,MarcoReflection,IRSalouiniBlockage}) along this line.
Considering a finite number of co-channel/interfering BSs, joint active/passive beamforming design with a cell-edge IRS is investigated for the users' weighted-sum-rate maximization\cite{IRSmulticellNallan} or minimum-rate maximization\cite{IRSmulticellXujie}, respectively.
In \cite{IRSmulticellCuiYing}, the authors consider the quasi-static phase-shift design of one IRS in the presence of one interfering BS, based on the statistical channel state information (CSI) assuming given BS/IRS locations.
In \cite{IRSdistributedMultiPairsShiYuanming}, the sum rate of multiple transmit-receive (Tx-Rx) pairs aided by multiple distributed IRSs at given locations is maximized.
However, the above works only consider a given number of BSs and IRSs at fixed locations, but do not investigate the impact of their spatial random locations on the performance of large-scale hybrid active/passive wireless networks.

Due to practical space constraints and heterogeneous/dense BS deployment, modern cellular networks typically exhibit an increasing degree of spatial irregularity, for which the conventional grid-based BS deployment models become no more suitable. As field trials are costly and system-level simulations are time-consuming, stochastic geometry has been extensively applied as a tractable analytical tool to model the spatial distribution of heterogeneously/densely/irregularly deployed wireless nodes, which provides meaningful performance lower bounds and scaling laws for practical wireless networks\cite{AndrewsCellular}\cite{SGexperiment}.
Based on stochastic geometry, the authors in \cite{MarcoReflection} model the IRSs by boolean line segments in a large-scale network and derive the probability that a given IRS is capable of providing an indirect path for a given Tx-Rx pair (i.e., the reflection probability).
The authors in \cite{IRSalouiniBlockage} further exploit the deployment of IRSs for providing indirect line-of-sight (LoS) paths for blocked links, thus improving the coverage probability in a large-scale network.
However, these two works do not consider the inter-cell interference and the small-scale fading effect.
To our best knowledge, the modeling of a general multi-cell hybrid wireless network aided by randomly located IRSs and the characterization of the distribution of the users' achievable signal-to-interference-plus-noise ratio (SINR) as well as the spatial throughput of the hybrid network have not been investigated yet in the literature.

Motivated by the above, in this paper, we model a hybrid active/passive wireless network under the general multi-cell setup and derive the distributions of the signal power, interference power, and thereby the users' achievable SINR in the network, with the ultimate goal of characterizing the \textit{spatial throughput} of the network, defined as the achievable rate per user equipment (UE) averaged over both the wireless channel fading and the random BS/IRS locations.
Thus, this work is a substantial extension of our prior work\cite{IRSlyuSingleCell} under the single-cell setup to the more general multi-cell setup.
We focus on the downlink communication from the BSs to the UEs while the proposed analytical framework can be similarly extended to the uplink communication, which is left for our future work.
Compared to other prior works on characterizing the performance of wireless networks with active BSs only (see, e.g., \cite{AndrewsCellular} and references therein), our analysis needs to derive the power distributions of both the signal and interference reflected by distributed IRSs in the network under spatially correlated channels, which exhibit channel hardening effects when the number of IRS elements becomes large.
Our main contributions are summarized as follows.

\begin{itemize}[leftmargin=0.14in]
	\item First, we model the random BS/IRS locations by independent homogeneous Poisson point processes (HPPPs) and propose a practical UE-to-IRS association rule when they are in close proximity. Then, for a typical BS-IRS-UE link with their given locations, we derive its channel power distribution in terms of the number of reflecting elements per IRS, denoted by $N$, based on which the mean channel power is shown to scale with $N$ in the order of $O(N^2)$ and $O(N)$ for the cases with reflect beamforming by the associated IRS and random scattering by non-associated IRSs, respectively.
	Furthermore, we define the network coverage probability and spatial throughput in terms of key system parameters including the BS/IRS densities and network loading factor.

	\item Next, we propose an analytical framework for the IRS-aided hybrid network based on stochastic geometry, and address its new challenges. In particular, for a typical UE 0 with randomly distributed IRSs nearby, its nearest IRS 0 (and hence the IRS 0-UE 0 distance $d_0$) typically has the dominant impact on the mean signal or interference power compared to other (farther) IRSs, under practical values of the IRS density.
Moreover, with reflect beamforming by the associated IRS, the signal link exhibits \textit{channel hardening} when $N$ becomes large, while the extent of channel hardening varies with the distance $d_0$, rendering it difficult to characterize the signal power distribution and thus the SINR distribution. To overcome this difficulty, we propose to approximate the conditional signal power distribution by the Gamma distribution, whose shape parameter $k_S$ specifies the extent of channel hardening conditioned on $d_0$.
	The conditional SINR distribution is then obtained in terms of the interference power Laplace transform and its derivatives up to integer-order $k_S$. Moreover, we propose an interpolation method for non-integer $k_S$, and apply the normal approximation of the signal power in the case with large $k_S$ in order to reduce the computational complexity.
	These new analytical methods jointly yield accurate and efficient characterization of the network SINR distribution and hence its spatial throughput.

	\item Finally, extensive numerical results are provided to validate our analytical results.
	It is found that increasing IRS density in a hybrid wireless network can \textit{significantly enhance the signal power but with only marginally increased interference}, thus greatly improving its throughput as compared to the traditional wireless network with active BSs only, especially when the BS density, network loading factor or $N$ is large.
	Moreover, it is unveiled that \textit{there exists an optimal IRS/BS density ratio} $\zeta^*$ for maximizing the spatial throughput of the new hybrid network under a given total deployment cost, where $\zeta^*$ is shown to increase with the BS/IRS cost ratio and the network loading factor,
	while the conventional network without IRS (i.e., zero IRS/BS density ratio), or the hybrid network with excessively large $\zeta$ (where the BS density is too low to provide enough signal power for effective IRS passive beamforming), is generally suboptimal in terms of throughput per unit cost.
	Furthermore, it is shown that the maximum spatial throughput of the hybrid network with the optimal IRS/BS density ratio $\zeta^*$ grows almost linearly with the total cost, thus providing a new and cost-effective approach to achieve \textit{sustainable capacity growth} for future wireless networks.
\end{itemize}

The rest of this paper is organized as follows.
The new model of the proposed hybrid wireless network is presented in Section \ref{SectionModel}.
The distributions/mean values of the signal and interference powers are then characterized in Section \ref{SectionSignal} and Section \ref{SectionInterference}, respectively.
Next, the SINR distribution and the network spatial throughput are obtained in Section \ref{SectionCharacterization}.
Numerical results are provided in Section \ref{SectionSimulation}.
Finally, we conclude the paper in Section \ref{SectionConclusion}.

\section{System Model}\label{SectionModel}

\begin{figure}
	\centering
	\includegraphics[width=1\linewidth,  trim=0 0 0 0,clip]{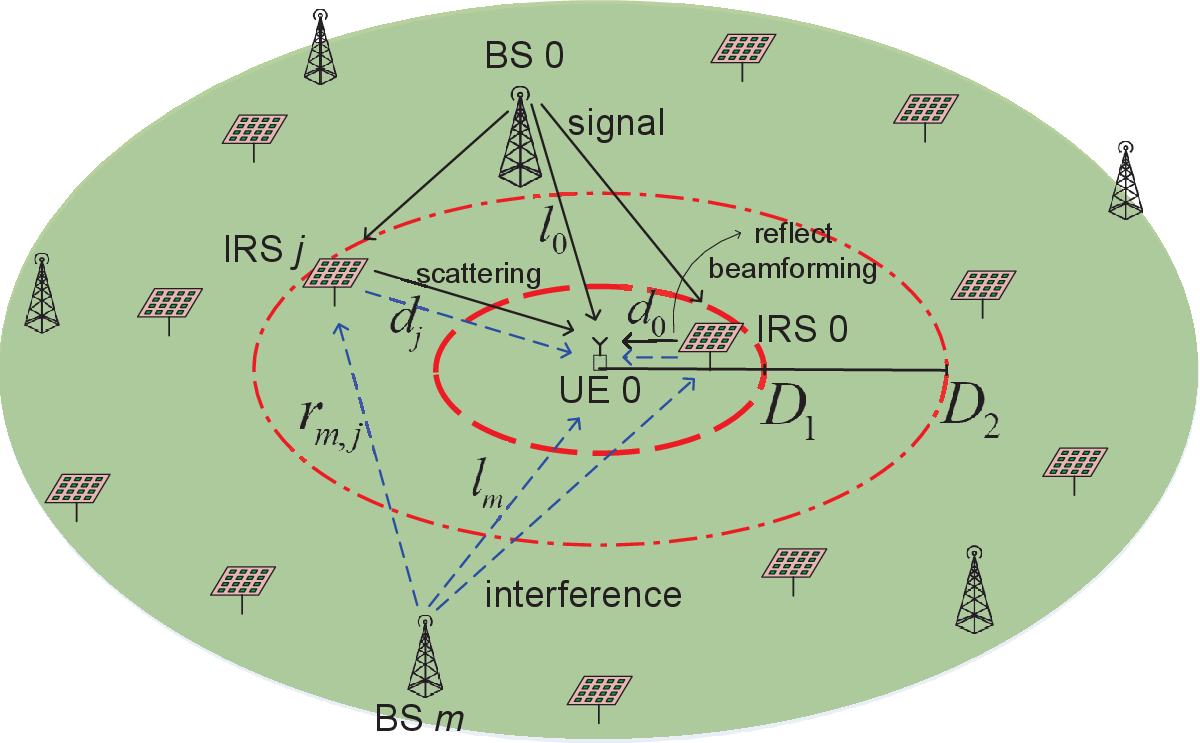}
	\caption{IRS-aided multi-cell wireless network in the downlink.\vspace{-2ex}}\label{IRS}
\end{figure}

In this paper, we consider an IRS-aided multi-cell wireless network shown in Fig. \ref{IRS}, and focus on the downlink communication from the BSs to UEs.
Assume that the BSs are of the same height equal to $H_\textrm{B}$ meters (m), while the BSs' horizontal locations are modeled by a 2-dimensional (2D) HPPP $\Lambda_\textrm{B}$ on the ground plane with given density $\lambda_\textrm{B}$ BSs/m$^2$.
Assume that the UE locations follow another independent HPPP on the ground plane,\footnote{The HPPP assumption implies the uniformly random UE distribution, which is more accurate for UEs of homogeneous distribution and/or higher mobility, but in general can serve as a good baseline to evaluate the network performance with heterogeneous user distribution/mobility in practice.} such that we can focus on one typical UE 0 in this HPPP to analyze the average UE performance without changing the location distribution of other UEs according to the Slivnyak's theorem\cite{AndrewsCellular}.
To facilitate our analysis, we consider the orthogonal frequency division multiple access (OFDMA) scheme and assume that the transmission bandwidth and each time slot are equally divided into orthogonal resource blocks (RBs), each randomly assigned by a BS to one of its served UEs, over which the channel is assumed to be frequency-flat and constant, while the channels may vary over different frequency bands or different time slots. 
We assume that the network has a homogeneous traffic load, where all BSs have a common loading factor $p$ $(0<p\le 1)$, i.e., each of the RBs is active with probability $p$ independently.
Consider one typical RB used by the typical UE 0 located at the origin, which is associated with its nearest BS 0 with distance $l_0$, as shown in Fig. \ref{IRS}. As a result, the BSs that transmit on the same RB form a thinned HPPP $\Lambda_\textrm{B}'$ with density $\lambda_\textrm{B}'\triangleq p\lambda_\textrm{B}$.

We consider that distributed IRSs are deployed to assist the BS-UE communications in the network. 
Assume that all IRSs are of the same height equal to $H_\textrm{I}$ m,\footnote{Our analytical framework is applicable to any given height of BSs/IRSs, which can also be extended to account for their random heights by employing the 3-dimensional point process where the height of each BS/IRS is randomly set within a certain range.} while the IRSs' horizontal locations are modeled by a 2D HPPP $\Lambda_\textrm{I}$ (independent of $\Lambda_\textrm{B}$) on the ground plane with given density $\lambda_\textrm{I}$ IRSs/m$^2$.
Denote the set of IRS horizontal locations as $\mathcal{W}\triangleq\{\bold w_j\in\mathbb{R}^2|j\in\Lambda_\textrm{I}\}$, where $\bold w_j$ is the 2D coordinate of an IRS $j\in\Lambda_\textrm{I}$.
Denote $d_j$ as the horizontal distance between UE 0 and IRS $j$, as shown in Fig. \ref{IRS}.
Since the IRS typically provides signal enhancement via reflect beamforming in a \textit{local} region \cite{IRSlyuSingleCell}, we consider the practical scenario where UE 0 is associated with its nearest IRS 0 for dedicated reflect beamforming, if IRS 0 is within a certain threshold distance $D_1$, i.e., $d_0\leq D_1$.\footnote{The threshold $D_1$ is practically set such that each IRS serves a finite number of UEs in its neighborhood only.} 
On the other hand, if there is no IRS within $D_1$, then UE 0 is served by BS 0 only without any associated IRS.\footnote{To implement this, each IRS controller can sense the nearby UEs and decide to associate with them or not based on their signal strengths, then send the associated UEs’ identification back to the BS (via a separate control link) for RB allocations. Note that such IRS-UE associations change only when the UEs move in/out from the coverage of each IRS, which usually happens not so frequently for typical IRS-aided scenarios (e.g., hotspot with local users).}
Moreover, for the purpose of exposition, we assume that each IRS is always on and reflects the received wave at all time, regardless of whether there is any UE associated with it.\footnote{The results in this paper can be extended to the general case where each IRS is independently on or off with a certain probability. Nevertheless, we consider that all IRSs are on to characterize the worst-case interference.}
As a result, UE 0 receives the reflected signal/interference by all IRSs (including its associated IRS 0 if any).
\rev{To model the randomly reflected signal/interference by non-serving IRSs accurately while maintaining analytical tractability, we adopt the \textbf{Approximation 1} that only the IRSs within a sufficiently large threshold distance $D_2$ ($D_2>D_1$) from UE 0, denoted by the set $\mathcal{J}\triangleq\{j\in\Lambda_\textrm{I}|d_j\leq D_2\}$, will contribute the signal/interference to it.}
Finally, to maximize the passive beamforming gain of the IRS to each served UE, we assume that its served UEs are assigned in \textit{orthogonal-time} RBs, i.e., time division multiple access (TDMA) or time sharing is adopted for the UEs served by the same IRS.\footnote{It is shown in \cite{IRSbeixiongNOMA} that for IRS-aided multiple access, the TDMA scheme is in general superior over the FDMA scheme due to the hardware limitation of IRS passive reflection, which can be made time-selective, but not frequency-selective\cite{QQirsMag}.}

\subsection{Channel Model}

Assume for simplicity that the BSs and UEs are each equipped with a single antenna, while each IRS has $N$ reflecting elements.
The baseband equivalent channels from BS $m$ to IRS $j$, from IRS $j$ to UE 0, and from BS $m$ to UE 0 are denoted by $\bold h_{\textrm{i},m}^{(j)}\triangleq[h_{\textrm{i},m,1}^{(j)},\cdots,h_{\textrm{i},m,N}^{(j)}]^T\in \mathbb{C}^{N\times 1}$, $\bold h_{\textrm{r}}^{(j)}\triangleq[h_{\textrm{r},1}^{(j)},\cdots,h_{\textrm{r},N}^{(j)}]^T\in \mathbb{C}^{N\times 1}$, and $h_{\textrm{d},m}\in \mathbb{C}$, respectively,\footnote{\rev{The subscripts ``i", ``r" and ``d" represent the BS-to-\textit{IRS} channel, IRS-\textit{reflected} channel (i.e., the IRS-to-UE channel), and \textit{direct} BS-to-UE channel, respectively.}} where $\mathbb{C}$ denotes the set of complex numbers and $[\cdot]^T$ denotes the matrix transpose.
Let $\boldsymbol\phi^{(j)}\triangleq [\phi_1^{(j)}, \cdots,\phi_N^{(j)}]$ and further denote $\boldsymbol\Phi^{(j)}\triangleq \diag\{ [e^{\bold i\phi_1^{(j)}}, \cdots,e^{\bold i\phi_N^{(j)}}]\}$ (with $\bold i$ denoting the imaginary unit) as the phase-shifting matrix of IRS $j$, where $\phi_n^{(j)}\in[0,2\pi)$ is the phase shift by element $n$ of the IRS on the incident signal,\footnote{In this paper, we assume (maximum) unit amplitude for each reflection coefficient to maximize the IRS beamforming gain to its served UE\cite{QQtwc}.} and $\diag\{\bold x\}$
denotes a diagonal matrix with each diagonal element being the corresponding element in $\bold x$. 
Each element of the IRS receives the superposed multi-path signals from the BS, and scatters the combined signal with adjustable phase as if from a single point source. Therefore, the cascaded BS-IRS-UE channel can be modeled as a concatenation of three components, namely, BS-IRS link, IRS reflecting with phase shifts, and IRS-UE link, given by\cite{QQtwc}
\begin{equation}\label{hirmAll}
h_{\textrm{ir},m}^{(j)}\triangleq [\bold h_{\textrm{i},m}^{(j)}]^T \boldsymbol\Phi^{(j)} \bold h_{\textrm{r}}^{(j)}=\sum\limits_{n=1}^N h_{\textrm{i},m,n}^{(j)} h_{\textrm{r},n}^{(j)} e^{\bold i\phi_n^{(j)}},m\in \Lambda_\textrm{B},
\end{equation}
where $h_{\textrm{ir},m,n}^{(j)}\triangleq h_{\textrm{i},m,n}^{(j)} h_{\textrm{r},n}^{(j)} e^{\bold i\phi_n^{(j)}}$ denotes the BS $m$-IRS $j$-UE 0 channel reflected by element $n$, $n=1,\cdots,N$.

For the BS-IRS, IRS-UE and BS-UE links, we assume a simplified fading channel model without shadowing, which consists of distance-dependent path loss with path-loss exponent $\alpha\geq 2$ and an additional random term $\xi$ accounting for small-scale fading.
Note that the shadowing effect can also be incorporated into the stochastic geometry based analytical framework by treating it as equivalent random perturbation in the locations of the BSs/IRSs (see Sections III.G and VI.A in \cite{AndrewsPrimer} and the references therein), which is ignored in this work for simplicity.
The channel power gain from BS $m$ to UE 0 is thus given by
\begin{equation}\label{gau}
|h_{\textrm{d},m}|^2\triangleq g_{\textrm{d},m}\xi_{\textrm{d},m}=\beta (l_m^2+H_\textrm{B}^2)^{-\alpha/2}\xi_{\textrm{d},m},
\end{equation}
where $g_{\textrm{d},m}$ is the average channel power gain, \rev{$\xi_{\textrm{d},m}$ accounts for channel fading,} $l_m$ denotes the BS-UE horizontal distance, and $\beta=(\frac{4\pi f_c}{c})^{-2}$ denotes the average channel power gain at a reference distance of 1 m based on the free-space path-loss model, with $f_c$ denoting the carrier frequency, and $c$ denoting the speed of light.
Similarly, the channel power gains from BS $m$ to the $n$-th element of IRS $j$, and from the latter to UE 0 are given by
\begin{equation}\label{gai}
|h_{\textrm{i},m,n}^{(j)}|^2\triangleq g_{\textrm{i},m}^{(j)}\xi_{\textrm{i},m,n}^{(j)}=\beta \big(r_{m,j}^2+(H_\textrm{B}-H_\textrm{I})^2\big)^{-\alpha/2}\xi_{\textrm{i},m,n}^{(j)},
\end{equation}
and
\begin{equation}\label{giu}
|h_{\textrm{r},n}^{(j)}|^2\triangleq g_{\textrm{r}}^{(j)}\xi_{\textrm{r},n}^{(j)}=\beta \big(d_j^2+H_\textrm{I}^2\big)^{-\alpha/2}\xi_{\textrm{r},n}^{(j)},
\end{equation}
where $g_{\textrm{i},m}^{(j)}$ and $g_{\textrm{r}}^{(j)}$ denote the average channel power gains,  \rev{$\xi_{\textrm{i},m,n}^{(j)}$ and $\xi_{\textrm{r},n}^{(j)}$ account for channel fading,} $r_{m,j}$ and $d_j$ denote the BS-IRS and IRS-UE horizontal distances,\footnote{For the purpose of exposition, we consider far-field propagation for all links, and accordingly assume $H_\textrm{B}\geq 1$ m and $H_\textrm{I}\geq 1$ m, which also avoid unbounded power gain when the horizontal distance $l_m$ or $d_j$ becomes zero.} respectively.

\subsection{BS-IRS-UE Channel Power Statistics}\label{Statistics}
In this subsection, we derive the BS-IRS-UE cascaded channel power statistics, which is new for the IRS-aided hybrid network and essential to our subsequent performance analysis for it.
Assume that the channels $h_{\textrm{d},m}$, $h_{\textrm{i},m,n}^{(j)}$ and $h_{\textrm{r},m,n}^{(j)}$, $m\in\Lambda_\textrm{B}$, $j\in\Lambda_\textrm{I}$, $n=1,\cdots,N$
are independent.
For the cascaded BS $m$-IRS $j$-UE 0 link in \eqref{hirmAll}, the channel reflected through each element $n$ is given by
\begin{align}\label{hirmn}
h_{\textrm{ir},m,n}^{(j)}&\triangleq h_{\textrm{i},m,n}^{(j)} h_{\textrm{r},n}^{(j)} e^{\bold i\phi_n^{(j)}}\notag\\
&=|h_{\textrm{i},m,n}^{(j)}| |h_{\textrm{r},n}^{(j)}| e^{\bold i\big(\phi_n^{(j)}+\angle h_{\textrm{i},m,n}^{(j)}+\angle h_{\textrm{r},n}^{(j)}\big)},
\end{align}
\rev{where the channel amplitude $|h_{\textrm{ir},m,n}^{(j)}|\triangleq|h_{\textrm{i},m,n}^{(j)}| |h_{\textrm{r},n}^{(j)}|$ resembles a \textit{doubly-faded} RV} while the channel phase $\angle h_{\textrm{ir},m,n}^{(j)}\triangleq \phi_n^{(j)}+\angle h_{\textrm{i},m,n}^{(j)}+\angle h_{\textrm{r},n}^{(j)}$ is adjustable via controlling the phase shift $\phi_n^{(j)}$ exerted by IRS $j$.

\rev{For simplicity, assume that $\xi_{\textrm{d},m}\stackrel{\textrm{dist.}}{=}\xi\sim \textrm{Exp}(1)$ is an exponential random variable (RV) with unit mean accounting for the small-scale Rayleigh fading.
	Therefore, the amplitude $|h_{\textrm{d},m}|$ follows the Rayleigh distribution\cite{forbes2011statistical} with scale parameter $\sqrt{g_{\textrm{d},m}/2}$, denoted by $\mathcal{R}\big(\sqrt{g_{\textrm{d},m}/2}\big)$, while $h_{\textrm{d},m}$ follows the circularly symmetric complex Gaussian (CSCG) distribution\cite{forbes2011statistical} with mean zero and covariance $g_{\textrm{d},m}$, denoted by $\mathcal{CN}(0,g_{\textrm{d},m})$.
	We also assume Rayleigh faded channel for the BS-IRS and IRS-UE links, i.e., $\xi_{\textrm{i},m,n}^{(j)}, \xi_{\textrm{r},n}^{(j)}\stackrel{\textrm{dist.}}{=}\xi\sim \textrm{Exp}(1)$, in order to investigate the worse-case propagation condition for the IRS and hence characterize the achievable performance lower bound for the IRS-aided hybrid network.\footnote{\rev{The proposed analytical method in this paper can be extended to other fading channel models such as Rician fading.}}
	Therefore, we have $|h_{\textrm{i},m,n}^{(j)}|\sim\mathcal{R}\big(\sqrt{g_{\textrm{i},m}^{(j)}/2}\big)$ and $|h_{\textrm{r},n}^{(j)}|\sim\mathcal{R}\big(\sqrt{g_{\textrm{r}}^{(j)}/2}\big)$.}

\rev{As a result, the channel amplitude $|h_{\textrm{ir},m,n}^{(j)}|$ in \eqref{hirmn} is a \textit{double-Rayleigh} RV.}
Note that for a double-Rayleigh distributed RV $Y=X_1X_2$ with independent $X_1\sim \mathcal{R}(\delta_1)$ and $X_2\sim \mathcal{R}(\delta_2)$, its mean and variance are respectively given by\cite{DoubleRayleigh}
\begin{equation}
\mathbb{E}\{Y\}\triangleq \pi\delta_1\delta_2/2,
\end{equation}
\begin{equation}
\var\{Y\}\triangleq 4\delta_1^2\delta_2^2(1-\pi^2/16).
\end{equation}
Therefore, the mean and variance of the channel amplitude $|h_{\textrm{ir},m,n}^{(j)}|$ are respectively given by
\begin{equation}\label{hirmean}
\mathbb{E}\big\{\big|h_{\textrm{ir},m,n}^{(j)}\big|\big\}\triangleq\frac{\pi}{4}\sqrt{g_{\textrm{i},m}^{(j)}g_{\textrm{r}}^{(j)}},
\end{equation}
\begin{equation}\label{hirvariance}
\var\big\{\big|h_{\textrm{ir},m,n}^{(j)}\big|\big\}\triangleq(1-\pi^2/16)g_{\textrm{i},m}^{(j)}g_{\textrm{r}}^{(j)}.
\end{equation}%

In the case where IRS 0 provides reflect beamforming service for the desired signal from BS 0 to UE 0 (i.e., $d_0\leq D_1$),
we assume that the cascaded channel phase $\angle\big(h_{\textrm{i},0,n}^{(0)} h_{\textrm{r},n}^{(0)}\big)$ for each reflected path $n=1,\cdots,N$ can be obtained via IRS-customized channel estimation methods (please refer to \cite{IRSyifeiOFDM}\cite{IRSzhengBeixiong} for more details).
As a result, IRS 0 can then adjust the phase shift $\boldsymbol\phi^{(0)}$ such that the $N$ reflected paths of the desired signal are of the same phase at UE 0's receiver by setting $\phi_n^{(0)}=-\angle\big(h_{\textrm{i},0,n}^{(0)}h_{\textrm{r},n}^{(0)}\big), n=1,\cdots,N$.\footnote{For the ease of practical implementation, we consider reflect beamforming for enhancing the desired signal power only, instead of nulling any co-channel interference.}
Therefore, the amplitude of the BS 0-IRS 0-UE 0 channel is given by
\begin{equation}\label{hReflectBeamforming}
|h_{\textrm{ir},0}^{(0)}|=|\bold h_{\textrm{i},0}^{(0)}|^T |\bold h_{\textrm{r}}^{(0)}|=\sum\limits_{n=1}^N \big|h_{\textrm{ir},0,n}^{(0)}\big|.
\end{equation}%
\rev{Furthermore, we adopt the \textbf{Approximation 2} that by the central limit theorem (CLT), given $N$ i.i.d. RVs $X_1,\cdots,X_N$ each with mean $\mu$ and variance $\omega^2$, the sum $Y=\sum_{n=1}^{N}X_n$ can be approximated by the normal distribution $\mathcal{N}(N\mu,N\omega^2)$ for sufficiently large $N$.}
As a result, we have
\newtheorem{prop}{Proposition}
\begin{prop}\label{Prop_Beamforming}
	With reflect beamforming by IRS 0, the BS 0-IRS 0-UE 0 channel amplitude in \eqref{hReflectBeamforming} for practically large $N$ can be approximated by the normal/Gaussian distribution, i.e.,
	\begin{align}\label{haiu0}
	|h_{\textrm{ir},0}^{(0)}|&\stackrel{\textrm{approx.}}{\sim}\mathcal{N}\bigg( N \mathbb{E}\big\{\big|h_{\textrm{ir},0,n}^{(0)}\big|\big\},N \var\big\{\big|h_{\textrm{ir},0,n}^{(0)}\big|\big\}\bigg)\notag\\
	&=\mathcal{N}\bigg(N\frac{\pi}{4}\sqrt{g_{\textrm{i},0}^{(0)} g_{\textrm{r}}^{(0)}}, N\big(1-\frac{\pi^2}{16}\big) g_{\textrm{i},0}^{(0)}g_{\textrm{r}}^{(0)}\bigg).
	\end{align}
\end{prop}
\textit{Proof:} \rev{Based on \textbf{Approximation 2} with large $N$,}\footnote{\rev{According to our simulations, when $N>25$, this approximation is already quite accurate for our considered setup.}} the BS 0-IRS 0-UE 0 channel amplitude in \eqref{hReflectBeamforming} is the sum of $N$ i.i.d. double-Rayleigh RVs $\big|h_{\textrm{ir},0,n}^{(0)}\big|$, $n=1,\cdots,N$, each with mean and variance given by \eqref{hirmean} and \eqref{hirvariance}, respectively. Therefore, Proposition \ref{Prop_Beamforming} is proved.$\blacksquare$

The average BS 0-IRS 0-UE 0 signal power is the second moment of $|h_{\textrm{ir},0}^{(0)}|$ and thus given by 
\begin{align}\label{gir0}
g_{\textrm{ir},0}^{(0)}&\triangleq \mathbb{E}\{|h_{\textrm{ir},0}^{(0)}|^2\}=\big(\mathbb{E}\{|h_{\textrm{ir},0}^{(0)}|\}\big)^2 + \var\{|h_{\textrm{ir},0}^{(0)}|\}\notag\\
&= \bigg[\frac{\pi^2}{16}N^2+\big(1-\frac{\pi^2}{16}\big)N\bigg]g_{\textrm{i},0}^{(0)}g_{\textrm{r}}^{(0)},
\end{align}%
which is proportional to the average channel power product $g_{\textrm{i},0}^{(0)}g_{\textrm{r}}^{(0)}$, with the beamforming gain coefficient $G_\textrm{bf}\triangleq \frac{\pi^2}{16}N^2+\big(1-\frac{\pi^2}{16}\big)N$ that grows with $N$ in the order of $O(N^2)$.

On the other hand, for any IRS $j$ that does not provide reflect beamforming for UE 0 (including IRS 0 if the distance $d_0>D_1$), it scatters the incoming signal from BS $m$ without passive beamforming, thus resulting in uniformly random channel phase $\angle h_{\textrm{ir},m,n}^{(j)}$ due to the uniformly random phases $\angle h_{\textrm{i},m,n}^{(j)}$ and $\angle h_{\textrm{r},n}^{(j)}$.
In this case, we have the following proposition.
\begin{prop}\label{Prop_Scattering}
	With random scattering by IRS $j$, the BS $m$-IRS $j$-UE 0 channel in \eqref{hirmAll} for practically large $N$ can be approximated by the CSCG distribution, i.e.,
	\begin{equation}\label{haium}
	h_{\textrm{ir},m}^{(j)}=\sum_{n=1}^N h_{\textrm{ir},m,n}^{(j)}\stackrel{\textrm{approx.}}{\sim} \mathcal{CN}\bigg(0,N g_{\textrm{i},m}^{(j)}g_{\textrm{r}}^{(j)}\bigg).
	\end{equation}
\end{prop}
\textit{Proof:} The channel $h_{\textrm{ir},m,n}^{(j)}$ reflected by each element $n$ has zero mean and independent in-phase and quadrature-phase components each with variance $\frac{1}{2}g_{\textrm{i},m}^{(j)}g_{\textrm{r}}^{(j)}$, respectively, with the detailed derivations given in Appendix \ref{AppendixCSCG}.
Moreover, since the channels $h_{\textrm{ir},m,n}^{(j)}$, $n=1,\cdots,N$ are i.i.d., \rev{based on \textbf{Approximation 2}}, the independent in-phase and quadrature-phase components of the combined channel $h_{\textrm{ir},m}^{(j)}=\sum_{n=1}^N h_{\textrm{ir},m,n}^{(j)}$ can be each approximated by an independent normal distribution $\mathcal{N}(0,\frac{1}{2}N g_{\textrm{i},m}^{(j)}g_{\textrm{r}}^{(j)})$ for practically large $N$.
As a result, the combined BS $m$-IRS $j$-UE 0 channel can be approximated by the CSCG distribution given by \eqref{haium}. Proposition \ref{Prop_Scattering} is thus proved.$\blacksquare$

Therefore, the average BS $m$-IRS $j$-UE 0 channel power is given by
\begin{equation}\label{gir}
g_{\textrm{ir},m}^{(j)}\triangleq \mathbb{E}\{|h_{\textrm{ir},m}^{(j)}|^2\}= N g_{\textrm{i},m}^{(j)}g_{\textrm{r}}^{(j)},
\end{equation}
which is proportional to the average channel power product $g_{\textrm{i},m}^{(j)}g_{\textrm{r}}^{(j)}$, with the scattering gain coefficient $G_\textrm{sc}\triangleq N$ that grows linearly with $N$.

\subsection{SINR, Coverage Probability, and Spatial Throughput}

Denote the downlink transmit power on each RB as $P_0$.
Then, the overall signal power (normalized by $P_0$) from BS 0 to UE 0 is given by 
\begin{equation}\label{SignalPower}
S\triangleq \bigg|h_{\textrm{d},0}+\sum_{j\in\mathcal{J}} h_{\textrm{ir},0}^{(j)}\bigg|^2,
\end{equation}
which accounts for the direct path and reflected paths via all IRSs $j\in\mathcal{J}$.
Further denote the total received interference power (normalized by $P_0$ as well) from all co-channel BSs $m\in \Lambda_\textrm{B}'\setminus\{0\}$ by
\begin{equation}\label{Idefinition}
	I\triangleq \sum_{m\in \Lambda_\textrm{B}'\setminus\{0\}} I_m= \sum_{m\in \Lambda_\textrm{B}'\setminus\{0\}}\bigg|h_{\textrm{d},m}+\sum_{j\in\mathcal{J}} h_{\textrm{ir},m}^{(j)}\bigg|^2.
\end{equation}
The received SINR at UE 0 is thus given by 
\begin{equation}\label{gamma}
\gamma\triangleq \frac{S}{I +W},
\end{equation}
where $W\triangleq \sigma^2/P_0$, and the receiver noise is assumed to be additive white Gaussian noise (AWGN) with power $\sigma^2$.

The corresponding achievable rate in bits/second/Hz (bps/Hz) is given by
\begin{equation}\label{Ck}
R\triangleq \log_2( 1+ \gamma).
\end{equation}
Note that the signal power $S$, interference power $I$, and thus SINR $\gamma$ and achievable rate $R$ are all RVs depending on the random channel fading as well as random BS/IRS locations.
An outage event occurs when the rate $R$ is lower than a minimum required target $\bar R$.
The coverage probability of the typical UE 0 is then defined as the average non-outage probability over the random channel fading and random BS/IRS locations, i.e.,
\begin{equation}
\textrm{P}_{\textrm{cov}}\triangleq \mathbb{P}\{R\geq\bar R\}=\mathbb{P}\{\gamma\geq\bar \gamma\}\triangleq 1-F_{\gamma}(\bar\gamma),
\end{equation}
where $\bar \gamma\triangleq 2^{\bar R}-1$ denotes the corresponding minimum required SINR, and  $F_{\gamma}(\cdot)$ is the cumulative distribution function (cdf)\footnote{The cdf of an RV $X$ is defined as $F_{X}(x)\triangleq \mathbb{P}\{X< x\}$.} of $\gamma$.
Accordingly, we can define the \textit{spatial throughput} of the network in bps/Hz/m$^2$ as
\begin{equation}\label{SpatialThroughput}
	\nu\triangleq \textrm{P}_{\textrm{cov}}\bar R \lambda_\textrm{B}'=\textrm{P}_{\textrm{cov}}\bar R p\lambda_\textrm{B}.
\end{equation}

In order to obtain the spatial throughput $\nu$, we need to characterize the cdf of the SINR $\gamma$ that depends on the distributions of the signal power $S$ and interference power $I$.
Note that the instantaneous $S$ and $I$ are independent due to their independent small-scale fading, while their large-scale statistics averaged over fading are dependent in general due to the common BS and IRS locations.
Specifically, under the distance-based association rule, the distribution of the BS 0-UE 0 link distance $l_0$ affects not only the mean signal power, but also the mean interference power since the interfering BSs are located at distances more than $l_0$ from UE 0.
Moreover, for the IRS-aided downlink communication, both the signal and interference are reflected by the same set ($\mathcal{J}$) of IRSs within $D_2$ of UE 0, and hence the path-losses of the BS-IRS and/or IRS-UE links are random but correlated in general.
Such correlation introduced by randomly distributed IRSs near UE 0 imposes new difficulty to the system-level performance analysis, which is challenging to deal with.
In addition, the IRS 0-UE 0 distance $d_0$ determines whether there is reflect beamforming provided by IRS 0 and thus the IRS 0-UE 0 channel power gain, which also has a significant impact on the system performance.

To tackle the above challenges, we decompose the performance analysis into three parts, by first characterizing the signal power and interference power distributions in Sections \ref{SectionSignal} and \ref{SectionInterference}, respectively, and then deriving the SINR distribution and hence the spatial throughput in Section \ref{SectionCharacterization}.
In particular, for the signal (or interference) power characterization, we first derive its \textit{conditional} distribution conditioned on the distances $l_0$ and $d_0$, based on which we are then able to obtain the SINR distribution. Moreover, we also characterize the \textit{unconditional} mean signal/interference power averaged over the channel fading and random BS/IRS locations, in order to reveal the impact of IRS on them, which helps illustrating their respective effects to the SINR distribution.

\section{Signal Power Distribution}\label{SectionSignal}

In this section, we characterize the conditional signal power distribution as well as the (unconditional) mean signal power.
Since in our considered IRS-aided hybrid network, the exact signal power distribution entails a more complicated form (as will be shown in Section \ref{SectionSignalAwith}) as compared to the conventional case without IRS, the well-known analytical method proposed in \cite{AndrewsCellular} for deriving the SINR distribution directly cannot be applied in our context.
Therefore, we propose to approximate the conditional signal power distribution by the Gamma distribution (which belongs to the exponential distribution family\cite{forbes2011statistical}) based on its first and second moments conditioned on $l_0$ and $d_0$, under three different cases based on $d_0$, i.e., with IRS reflect beamforming ($d_0\leq D_1$), with IRS scattering only ($D_1<d_0\leq D_2$), and without any nearby IRS ($d_0>D_2$).
Such Gamma approximation incorporates the BS-IRS-UE channel power statistics derived in Section \ref{Statistics} as well as the impact of randomly distributed IRSs, which is tailored to the new IRS-aided hybrid network and thus differs from that for the conventional network without IRS.
Moreover, for the case with IRS reflect beamforming, we characterize the impact of $d_0$ on the mean signal power and the $d_0$-dependent channel hardening effect to draw useful insights.
Finally, the mean signal power is obtained by integrating the conditional mean signal power over the distributions of $l_0$ and $d_0$.

Note that the first and second moments of the signal power $S$ depend on the average channel power gains $g_{\textrm{d},0}$ (BS 0-UE 0 link), $g_{\textrm{i},0}^{(j)}$ (BS 0-IRS $j$ link) and $g_{\textrm{r}}^{(j)}$ (IRS $j$-UE 0 link), which further depend on the corresponding horizontal link distance $l_0$, $r_{0,j}$ and $d_j$ that are related by the cosine law, i.e., $r_{0,j}^2=l_0^2+d_j^2-2l_0 d_j\cos\varphi$, where $\varphi$ is the BS 0-UE 0-IRS $j$ angle projected on the ground plane.
\rev{The exact expressions for the moments of $S$ conditioned on $l_0$ and $d_0$, though not expressible in closed-form, can be obtained by numerical integrals over $\varphi$ and/or $d_j$ (as illustrated in Appendix \ref{AppendixMoments}). Nevertheless, for simplicity in this section, we apply the \textbf{Approximation 3} that $r_{0,j}\approx l_0$ and hence $g_{\textrm{i},0}^{(j)}\approx g_{\textrm{d},0}$, $j\in\mathcal{J}$, in order to obtain closed-form approximations for the moments of $S|_{l_0,d_0}$. Such approximation is reasonable since the IRSs considered here are in the local region of UE 0 (and hence with small $d_j$), which is also verified by our numerical examples later in Section \ref{SectionSimulation} where the analytical results are shown to match well with the Monte Carlo (MC) simulation results.}

\subsection{The Case with IRS Reflect Beamforming}\label{SectionSignalAwith}

\subsubsection{Gamma Approximation with Moment Matching}
First, consider the case with $d_0\leq D_1$, where IRS 0 provides reflect beamforming for the desired signal from BS 0 to UE 0, while all other IRSs $j\in\mathcal{J}\setminus\{0\}$ randomly scatter signals.
We assume that the BS 0-UE 0 direct channel phase $\angle h_{\textrm{d},0}$ is known and IRS 0 can perform a common phase-shift such that $h_{\textrm{ir},0}^{(0)}$ and $h_{\textrm{d},0}$ are co-phased and hence coherently combined at UE 0\cite{QQtwc}.
On the other hand, other IRSs $j\in\mathcal{J}\setminus\{0\}$ randomly scatter signals, resulting in random phase $\angle h_{\textrm{ir},0}^{(j)}$ compared to $\angle h_{\textrm{d},0}$.

Denote $h_1\triangleq h_{\textrm{d},0}+ h_{\textrm{ir},0}^{(0)}=(|h_{\textrm{d},0}|+ |h_{\textrm{ir},0}^{(0)}|)e^{\bold i\angle h_{\textrm{d},0}}$ as the combined channel of the direct BS 0-UE 0 path and the BS 0-IRS 0-UE 0 path, and denote $h_2\triangleq \sum_{j\in\mathcal{J}\setminus\{0\}} h_{\textrm{ir},0}^{(j)}$ as the combined channel of the BS 0-IRS $j$-UE 0 paths for all other IRSs $j\in\mathcal{J}$ except IRS 0.
The overall signal power $S$ in \eqref{SignalPower} is then given by
\begin{equation}\label{Sexpand}
S\triangleq |h_1+h_2|^2=|h_1|^2+h_1^*h_2+h_1h_2^*+|h_2|^2,
\end{equation}
where $^*$ denotes the complex conjugate.
Since it is difficult to obtain the exact distribution of $S$, we propose to approximate it by the Gamma distribution based on its first and second moments conditioned on $l_0$ and $d_0$.

Based on \eqref{Sexpand},
the first two moments of $S$ are determined jointly by the first two moments of $h_1$, $h_2$, $|h_1|^2$ and $|h_2|^2$, which are derived in Appendix \ref{AppendixMoments}.
As a result, conditioned on $l_0$ and $d_0$, the first two moments of $S$ are respectively given by
	\begin{align}
	&\mathbb{E}\{S\}|_{l_0,d_0}=\mathbb{E}\{|h_1|^2+h_1^*h_2+h_1h_2^*+|h_2|^2\}|_{l_0,d_0}\notag\\
	&=\mathbb{E}\{|h_1|^2\}|_{l_0,d_0}+2 \mathbb{E}\{h_1\}|_{l_0,d_0}\mathbb{E}\{h_2\}|_{l_0,d_0}+\mathbb{E}\{|h_2|^2\}|_{l_0,d_0}\notag\\
	&=\mathbb{E}\{|h_1|^2\}|_{l_0,d_0}+\mathbb{E}\{|h_2|^2\}|_{l_0,d_0}\notag\\
	&\stackrel{(a)}{\approx} g_{\textrm{d}}(l_0)\bigg(1+G_\textrm{bf} g_{\textrm{r}}(d_0)+N\frac{\pi}{4}\sqrt{\pi g_{\textrm{r}}(d_0)}+N E_\textrm{I1}(d_0)\bigg),\label{meanSwithBF}
	\end{align}
	\begin{align}
	\mathbb{E}\{S^2\}|_{l_0,d_0}=&\mathbb{E}\big\{\big(|h_1|^2+h_1^*h_2+h_1h_2^*+|h_2|^2\big)^2\big\}|_{l_0,d_0}\notag\\
	=&\mathbb{E}\{|h_1|^4\}|_{l_0,d_0}+\mathbb{E}\{|h_2|^4\}|_{l_0,d_0}\notag\\
	&+4\mathbb{E}\{|h_1|^2\}|_{l_0,d_0}\mathbb{E}\{|h_2|^2\}|_{l_0,d_0},
	\end{align}%
\rev{where the approximation in $(a)$ is due to \textbf{Approximation 3};} $g_{\textrm{d}}(l)$ and $g_{\textrm{r}}(d)$ represent the average powers of the BS-UE direct link and the IRS-UE link, respectively, i.e.,
\begin{equation}\label{gd}
g_{\textrm{d}}(l)\triangleq \beta (l^2+H_\textrm{B}^2)^{-\alpha/2},
\end{equation}
\begin{equation}\label{gr}
g_{\textrm{r}}(d)\triangleq \beta (d^2+H_\textrm{I}^2)^{-\alpha/2},
\end{equation}
and $N\cdot E_\textrm{I1}(d_0)$ represents the expectation of the sum of IRS-UE channel powers from all IRSs within horizontal distance $(d_0,D_2]$ from UE 0, with $E_\textrm{I1}(d_0)$ given by
\begin{align}
&E_\textrm{I1}(d_0)\triangleq \mathbb{E}\bigg\{\sum_{d_0<d_j\leq D_2} g_{\textrm{r}}(d_j)\bigg\}=2\pi\lambda_\textrm{I}\int_{d=d_0}^{D_2}g_{\textrm{r}}(d)d\diff d\notag\\
&=\frac{2\pi\lambda_\textrm{I}\beta}{\alpha-2}\big[(d_0^2+H_\textrm{I}^2)^{1-\frac{\alpha}{2}}-(D_2^2+H_\textrm{I}^2)^{1-\frac{\alpha}{2}}\big].\label{EI1}
\end{align}

In the above, we have obtained closed-form expressions for the first and second moments of $S$ conditioned on the distances $l_0$ and $d_0$, and thereby the corresponding variance $\var\{S\}|_{l_0,d_0}\triangleq \mathbb{E}\{S^2\}|_{l_0,d_0}-(\mathbb{E}\{S\}|_{l_0,d_0})^2$.
As a result, for the case with IRS reflect beamforming, the Gamma distribution $\Gamma[k_\textrm{bf},\theta_\textrm{bf}]$ with the same first and second order moments as $S$ has the shape parameter $k_\textrm{bf}\triangleq (\mathbb{E}\{S\}|_{l_0,d_0})^2/\var\{S\}|_{l_0,d_0}$ and scale parameter $\theta_\textrm{bf}\triangleq \var\{S\}|_{l_0,d_0}/\mathbb{E}\{S\}|_{l_0,d_0}$.

\subsubsection{Impact of IRS 0-UE 0 Distance $d_0$}\label{SectionSignalWith_d0}
Based on \eqref{meanSwithBF}, the conditional mean signal power $\mathbb{E}\{S\}|_{l_0,d_0}$ is the product of two factors, namely $g_{\textrm{d}}(l_0)$ that depends on the location of BS 0, and $\kappa_\textrm{bf}(d_0)$ which depends on the location of IRS 0 and is given by
\begin{equation}\label{kappabf}
\kappa_\textrm{bf}(d_0)\triangleq 1+G_\textrm{bf} g_{\textrm{r}}(d_0)+N\frac{\pi}{4}\sqrt{\pi g_{\textrm{r}}(d_0)}+N E_\textrm{I1}(d_0).
\end{equation}
Note that in \eqref{kappabf} the dominant term is $G_\textrm{bf} g_{\textrm{r}}(d_0)$ that scales in $O(N^2)$ or $O\big(d_0^{-\frac{\alpha}{2}}\big)$ when $d_0$ is sufficiently small.
More specifically, for large $d_0$, the IRS-provided power gain $G_\textrm{bf} g_{\textrm{r}}(d_0)\ll 1$ and hence $\kappa_\textrm{bf}(d_0)\approx 1$.
On the other hand, as $d_0$ decreases, the IRS-provided power gain becomes higher and helps increase the mean signal power substantially, especially when $N$ is large.

On the other hand, the signal power variance $\var\{S\}|_{l_0,d_0}$ can be verified to scale in $O(N^3)$ or $O\big(d_0^{-\alpha}\big)$ when $d_0$ is sufficiently small.
As a result, it can be verified that the shape parameter $k_\textrm{bf}\triangleq (\mathbb{E}\{S\}|_{l_0,d_0})^2/\var\{S\}|_{l_0,d_0}$ scales in $O(N)$ when $d_0$ is sufficiently small.
Since $k_\textrm{bf}$ represents the square of the mean-to-standard deviation ratio, the increasing $k_\textrm{bf}$ with decreasing $d_0$ implies an increasingly more prominent ``channel hardening" effect since the mean signal power increases faster than its standard deviation as $N$ increases, thus leading to relatively less variations around the mean value. Such channel hardening results in a nearly deterministic channel, which improves the reliability with lower outage and also alleviates the need for frequent channel estimation.

The above results show that with any given $N$, IRS can provide more significant and reliable power gains for UEs that are closer to it, i.e., with smaller horizontal distance $d_0$ from it.
On the other hand, when the distance $d_0$ is small, increasing $N$ leads to increased mean signal power and also channel hardening effect, both of which are beneficial for the system performance.

\subsection{The Case without IRS Reflect Beamforming}\label{SectionSignalAwithout}
\subsubsection{Gamma Approximation with Moment Matching}
Next, consider the case with $D_1<d_0\leq D_2$, where IRS 0 is outside the association range $D_1$ of UE 0 and hence just randomly scatters its received signal, like all the other IRSs in $\mathcal{J}$.
In this case, we derive the first and second moments of the signal power $S$ conditioned on $l_0$ only, by averaging over the random locations of all IRSs outside $D_1$ of UE 0.
Specifically, for all IRSs $j\in\mathcal{J}$, the BS $0$-IRS $j$-UE 0 channel follows \eqref{haium} and is given by
\begin{equation}
h_{\textrm{ir},0}^{(j)}\stackrel{\textrm{approx.}}{\sim} \mathcal{CN}\big(0,N g_{\textrm{i},0}^{(j)}g_{\textrm{r}}^{(j)}\big).
\end{equation}
As a result, the composite signal channel $h_{\textrm{d},0}+\sum_{j\in\mathcal{J}} h_{\textrm{ir},0}^{(j)}$ from BS 0 is the sum of independent CSCG RVs, and hence is still CSCG distributed.
Therefore, the composite signal power $S$ follows the exponential distribution and is given by
\begin{equation}\label{Swithout}
S\triangleq \bar{S} \xi_0= \bigg(g_{\textrm{d},0}+N\sum_{j\in\mathcal{J}} g_{\textrm{i},0}^{(j)}g_{\textrm{r}}^{(j)}\bigg)\xi_0,
\end{equation}
where $\xi_0\stackrel{\textrm{dist.}}{=}\xi\sim \textrm{Exp}(1)$ and $\bar S$ is the mean signal power under given BS/IRS locations.
\rev{Based on \textbf{Approximation 3} that $g_{\textrm{i},0}^{(j)}\approx g_{\textrm{d},0}=g_{\textrm{d}}(l_0)$,} we hence have
\begin{equation}\label{barSapprox}
\bar S\approx g_{\textrm{d}}(l_0)\bigg[1 + N\sum_{j\in\mathcal{J}}g_{\textrm{r}}(d_j)\bigg].
\end{equation}
As a result, the conditional mean signal power is given by
\begin{equation}\label{meanSwithoutBF}
\mathbb{E}\{S\}|_{l_0}=\mathbb{E}\{\xi_0\}\mathbb{E}\{\bar S\}|_{l_0}\approx g_{\textrm{d}}(l_0)\big[1+ N E_\textrm{I1}(D_1)\big],
\end{equation}
where $E_\textrm{I1}(D_1)=E_\textrm{I1}(d_0)\big|_{d_0=D_1}$ based on \eqref{EI1}.
Similarly, the conditional second moment of $S$ is given by
\begin{align}
&\mathbb{E}\{S^2\}|_{l_0}=\mathbb{E}\{\xi_0^2\}\mathbb{E}\{\bar S^2\}|_{l_0}\notag\\
&\approx 2[g_{\textrm{d}}(l_0)]^2\bigg[1+2 N E_\textrm{I1}(D_1)+N^2 E_\textrm{I3}(D_1)\bigg],
\end{align}
where $E_\textrm{I3}(D_1)=E_\textrm{I3}(d_0)\big|_{d_0=D_1}$ and $E_\textrm{I3}(d_0)$ is given by
\begin{equation}\label{EI3}
E_\textrm{I3}(d_0)\triangleq \mathbb{E}\bigg\{\bigg(\sum_{d_0<d_j\leq D_2} g_{\textrm{r}}(d_j)\bigg)^2\bigg\}=\big(E_\textrm{I1}(d_0)\big)^2+E_\textrm{I2}(d_0),
\end{equation}
with $E_\textrm{I2}(d_0)$ defined as
\begin{align}
&E_\textrm{I2}(d_0)\triangleq \mathbb{E}\bigg\{\sum_{d_0<d_j\leq D_2} \big[g_{\textrm{r}}(d_j)\big]^2\bigg\}\notag\\
&=\frac{\pi\lambda_\textrm{I}\beta^2}{\alpha-1}\big[(d_0^2+H_\textrm{I}^2)^{1-\alpha}-(D_2^2+H_\textrm{I}^2)^{1-\alpha}\big].
\end{align}

Since we have obtained closed-form expressions for the first and second moments of $S$ conditioned on $l_0$, we can also obtain the corresponding variance $\var\{S\}|_{l_0}\triangleq \mathbb{E}\{S^2\}|_{l_0}-(\mathbb{E}\{S\}|_{l_0})^2$.
As a result, for the case with IRS scattering only, the Gamma distribution $\Gamma[k_\textrm{sc},\theta_\textrm{sc}]$ with the same first and second order moments as $S$ has the shape parameter $k_\textrm{sc}\triangleq (\mathbb{E}\{S\}|_{l_0})^2/\var\{S\}|_{l_0}$ and scale parameter $\theta_\textrm{sc}\triangleq \var\{S\}|_{l_0}/\mathbb{E}\{S\}|_{l_0}$.
Finally, it can be verified that the shape parameter $k_\textrm{sc}$ in this case does not scale with $N$, thus indicating no IRS-induced channel hardening effect in contrast to the case with IRS reflect beamforming shown in the previous subsection.

Finally, for the case without any nearby IRS (i.e., $d_0>D_2$), UE 0 is served by BS 0 directly, and hence the signal power $S=|h_{\textrm{d},0}|^2$ follows the exponential distribution with mean $g_{\textrm{d},0}=g_{\textrm{d}}(l_0)$, which is a special case of the Gamma distribution, i.e., $S\sim \Gamma[k_\textrm{wo},\theta_\textrm{wo}]$ with shape parameter $k_\textrm{wo}=1$ and scale parameter $\theta_\textrm{wo}=g_{\textrm{d}}(l_0)$.
In summary, we have the following proposition.
\begin{prop}\label{Prop_GammaApprox}
	\rev{Based on \textbf{Approximations 1}, \textbf{2} and \textbf{3},} the signal power distribution conditioned on $l_0$ and $d_0$ can be approximated by
	\begin{equation}\label{Sgamma}
	S|_{l_0,d_0}\stackrel{\textrm{approx.}}{\sim} \Gamma[k_S,\theta_S]= \begin{cases}
	\Gamma[k_\textrm{bf},\theta_\textrm{bf}], & \textrm{if} \ d_0\leq D_1;\\
	\Gamma[k_\textrm{sc},\theta_\textrm{sc}], & \textrm{if} \ D_1<d_0\leq D_2;\\
	\Gamma[k_\textrm{wo},\theta_\textrm{wo}], & \textrm{otherwise,}
	\end{cases}
	\end{equation}
	which has the same first and second moments of $S|_{l_0,d_0}$.
\end{prop}
\textit{Proof:} For the three different cases based on $d_0$ (i.e., with IRS reflect beamforming, with IRS scattering only, and without nearby IRS), we have respectively approximated $S|_{l_0,d_0}$ by the Gamma distribution with the same first and second moments. Therefore, Proposition \ref{Prop_GammaApprox} follows.$\blacksquare$

Note that the above signal power distribution as a function of $d_0$ is generally not continuous at the boundary point $D_1$ (or $D_2$), whereas the discontinuity gap is smaller when $D_1$ (or $D_2$) is larger.

\subsubsection{Impact of the Nearest IRS 0 versus Other IRSs}\label{SectionSignalWithout_d0}

To see the respective impact of the nearest IRS 0 and other IRSs in $\mathcal{J}$, for the case with IRS scattering only, we first express the mean signal power in \eqref{barSapprox} under given BS/IRS locations as
\begin{equation}
\bar S\approx g_{\textrm{d}}(l_0)\bigg[1 +N g_{\textrm{r}}(d_0) +N \sum_{j\in\mathcal{J}\setminus\{0\}}g_{\textrm{r}}(d_j)\bigg].
\end{equation}
Then, the mean signal power conditioned on $l_0$ and $d_0$ can be obtained as
\begin{equation}
\mathbb{E}\{S\}|_{l_0,d_0}\approx g_{\textrm{d}}(l_0) \kappa_\textrm{sc}(d_0),
\end{equation}
which is the product of two factors, namely $g_{\textrm{d}}(l_0)$ that depends on the location of BS 0, and $\kappa_\textrm{sc}(d_0)$ that depends on the location of IRS 0 and is given by
\begin{equation}\label{kappasc}
\kappa_\textrm{sc}(d_0)\triangleq 1+N g_{\textrm{r}}(d_0) +N E_\textrm{I1}(d_0).
\end{equation}

Note that the term $N E_\textrm{I1}(d_0)$ in \eqref{kappasc} represents the power gain brought by other IRSs in $\mathcal{J}$ on the mean signal power, which is typically much smaller than $N g_{\textrm{r}}(d_0)$ that represents the power gain brought by the nearest IRS 0, for practical values of the IRS density $\lambda_\textrm{I}$.
Similar observation can be made on \eqref{kappabf} for the case with IRS reflect beamforming.
\rev{As a result, among all IRSs in $\mathcal{J}$, the nearest IRS 0 (and hence the distance $d_0$) typically has the dominant impact on the mean signal power compared to other IRSs. 
Therefore, we choose to characterize the signal power distribution (as well as the distributions of the interference power and SINR in the sequel) conditioned on $d_0$, while representing the power gain brought by other IRSs by its mean value $N E_\textrm{I1}(d_0)$.}
In other words, the IRS density $\lambda_\textrm{I}$ impacts the system performance mainly via the distribution of the IRS 0-UE 0 distance $d_0$, whose probability density function (pdf) is given by
\begin{equation}
f_{d_0}(d_0)\triangleq 2\pi\lambda_\textrm{I} d_0 e^{-\lambda_\textrm{I}\pi d_0^2},
\end{equation}
where larger $\lambda_\textrm{I}$ leads to smaller $d_0$ on average and hence larger IRS-provided signal power gain.

\subsection{Mean Signal Power}\label{SectionSignalMean}

Last, we derive the mean signal power conditioned on $d_0$ for the above-mentioned three cases, respectively, which is then weighted by their probabilities of occurrence to obtain the unconditional mean signal power.

For the case with IRS reflect beamforming (i.e., $d_0\leq D_1$), the mean signal power conditioned on $d_0$ can be obtained by integrating over $l_0$ in \eqref{meanSwithBF}, which is given by
\begin{equation}\label{meanSwithBFd0}
\mathbb{E}\{S\}|_{d_0}\approx \kappa_\textrm{bf}(d_0) E_\textrm{B0},
\end{equation}
where $E_\textrm{B0}$ represents the expectation of the BS 0-UE 0 direct channel power, given by
\begin{align}\label{EB0}
&E_\textrm{B0}\triangleq \mathbb{E}\big\{g_{\textrm{d}}(l_0)\big\}=\int_{l_0=0}^\infty g_{\textrm{d}}(l_0)f_{l_0}(l_0)\diff l_0\notag\\
&=\beta\lambda_\textrm{B}\pi H_\textrm{B}^{2-\alpha} e^{\lambda_\textrm{B}\pi H_\textrm{B}^2} E_{\frac{\alpha}{2}}(\lambda_\textrm{B}\pi H_\textrm{B}^2),
\end{align}
where $f_{l_0}(l_0)\triangleq 2\pi\lambda_\textrm{B} l_0 e^{-\lambda_\textrm{B}\pi l_0^2}$ is the pdf of the BS 0-UE 0 distance $l_0$, and $E_{\frac{\alpha}{2}}(\cdot)$ is the exponential integral function\cite{NISTfunctions} with parameter $\frac{\alpha}{2}$, which is available in MATLAB.

Similarly, for the case with IRS scattering only (i.e., $D_1<d_0\leq D_2$), we have
\begin{equation}\label{meanSwithoutBFd0}
\mathbb{E}\{S\}|_{d_0}\approx \kappa_\textrm{sc}(d_0) E_\textrm{B0},
\end{equation}
with the occurrence probability given by
\begin{equation}
\textrm{P}_{\textrm{sc}}\triangleq e^{-\lambda_\textrm{I}\pi D_1^2}-e^{-\lambda_\textrm{I}\pi D_2^2}.
\end{equation}
Moreover, for the case without nearby IRS (i.e., $d_0> D_2$), we have $\mathbb{E}\{S\}|_{d_0}\approx E_\textrm{B0}$, with the occurrence probability $\textrm{P}_{\textrm{wo}}\triangleq e^{-\lambda_\textrm{I}\pi D_2^2}$.
Finally, we integrate $\mathbb{E}\{S\}|_{d_0}$ over $d_0$ to obtain the unconditional mean signal power $\mathbb{E}\{S\}$, i.e.,
\begin{small}
\begin{align}
&\mathbb{E}\{S\}=\int_{d_0=0}^\infty \mathbb{E}\{S\}|_{d_0} f_{d_0}(d_0) \diff d_0 \notag\\
&\approx E_\textrm{B0}\bigg[\int_{0}^{D_1} \kappa_\textrm{bf}(d_0) f_{d_0}(d_0)\diff d_0 + \int_{D_1}^{D_2} \kappa_\textrm{sc}(d_0) f_{d_0}(d_0)\diff d_0+\textrm{P}_{\textrm{wo}}\bigg]\notag\\
&=E_\textrm{B0}\bigg[\int_{d_0=0}^{D_1} \kappa_\textrm{bf}(d_0) f_{d_0}(d_0)\diff d_0 + \textrm{P}_{\textrm{sc}} E_\textrm{I1}(D_1) +\textrm{P}_{\textrm{wo}}\bigg].\label{Sint}
\end{align}%
\end{small}

\section{Interference Power Distribution}\label{SectionInterference}

In this section, we first characterize the interference power distribution due to random channel fading, under given BS/IRS locations. 
Based on this result, we then characterize the interference power distribution conditioned on the given distances $l_0$ and $d_0$, by deriving its Laplace transform\cite{NISTfunctions} and hence cdf.
Finally, we derive the mean interference power in the network.

\rev{Note that for analytical tractability in this section, we apply the \textbf{Approximation 4} that $r_{m,j}\approx l_m$ and hence $g_{\textrm{i},m}^{(j)}\approx g_{\textrm{d},m}$, $j\in\mathcal{J}$, which can be similarly justified as in the second paragraph of Section \ref{SectionSignal}.
Moreover, the distance $l_m$ is from the non-serving BSs $m\neq 0$ in other cells, which is at least larger than the distance $l_0$ from the serving BS 0 according to our assumed distance-based user-BS association. Therefore, the approximation of $r_{m,j}\approx l_m$ is more accurate in this case since $d_j$ is much smaller than $l_m$.}

\subsection{Interference Power Distribution Given BS/IRS Locations}\label{SectionScatter}

Based on \eqref{haium},
the BS $m$-IRS $j$-UE 0 channel can be approximated by the CSCG distribution $\mathcal{CN}\big(0,N g_{\textrm{i},m}^{(j)}g_{\textrm{r}}^{(j)}\big)$.
As a result, the composite interference channel $h_{\textrm{d},m}+\sum_{j\in\mathcal{J}} h_{\textrm{ir},m}^{(j)}$ from BS $m\in \Lambda_\textrm{B}'\setminus\{0\}$ is the sum of independent CSCG RVs, and hence is CSCG distributed with mean zero and covariance $\mathbb{E}\{|h_{\textrm{d},m}|^2\}+\sum_{j\in\mathcal{J}}\mathbb{E}\{|h_{\textrm{ir},m}^{(j)}|^2\}$.
Therefore, the composite interference power $I_m\triangleq \big|h_{\textrm{d},m}+\sum_{j\in\mathcal{J}} h_{\textrm{ir},m}^{(j)}\big|^2$ follows the exponential distribution and is given by
\begin{equation}\label{Im}
I_m\triangleq \bar I_m \xi_m= \bigg(g_{\textrm{d},m}+N\sum_{j\in\mathcal{J}} g_{\textrm{i},m}^{(j)}g_{\textrm{r}}^{(j)}\bigg)\xi_m,
\end{equation}
where $\xi_m\stackrel{\textrm{dist.}}{=}\xi\sim \textrm{Exp}(1)$ and $\bar I_m$ is the average interference power.
Therefore, the total interference power $I$ under given BS/IRS locations is the sum of independent but not identically distributed exponential RVs $I_m, m\in \Lambda_\textrm{B}'\setminus\{0\}$, and thus follows the generalized Erlang distribution\cite{forbes2011statistical}.

Note that the mean interference power under given BS/IRS locations is given by
	\begin{align}\label{barI}
	\bar I&\triangleq  \sum_{m\in \Lambda_\textrm{B}'\setminus\{0\}} \bar I_m\notag\\
	&= \bigg(\sum_{m\in \Lambda_\textrm{B}'\setminus\{0\}} g_{\textrm{d},m}\bigg)+ N\sum_{j\in\mathcal{J}}\bigg( g_{\textrm{r}}^{(j)}  \sum_{m\in \Lambda_\textrm{B}'\setminus\{0\}} g_{\textrm{i},m}^{(j)}\bigg).
	\end{align}%
\rev{Based on the definitions of $g_{\textrm{d}}(l)$ and $g_{\textrm{r}}(d)$ in \eqref{gd} and \eqref{gr}, as well as \textbf{Approximation 4} that $g_{\textrm{i},m}^{(j)}\approx g_{\textrm{d},m}=g_{\textrm{d}}(l_m)$ for $j\in\mathcal{J}$,} we have $\bar I_m\approx \eta g_{\textrm{d}}(l_m)$ and hence
\begin{equation}\label{barIapprox}
\bar I\approx \eta \sum_{m\in \Lambda_\textrm{B}'\setminus\{0\}} g_{\textrm{d}}(l_m),
\end{equation}%
where \rev{$\eta\triangleq 1+ N\sum_{j\in\mathcal{J}} g_{\textrm{r}}(d_j)$ is the relative power gain of all paths (including the scattering paths from all IRSs in $\mathcal{J}$) over the BS-UE direct path.
For each IRS $j\in\mathcal{J}$, its contributed term $N g_{\textrm{r}}(d_j)$ decreases with the IRS-UE distance $d_j$ in the order of $O(d_j^{-\alpha})$, which decays quickly and becomes negligible beyond a certain distance, thus justifying \textbf{Approximation 1}.}

\subsection{Conditional Laplace Transform and Cdf}

For complete characterization of the interference power distribution conditioned on $l_0$ and $d_0$, we derive its Laplace transform in the following.
From \eqref{Idefinition}, \eqref{Im} and \eqref{barIapprox}, we have
\begin{equation}\label{Iapprox}
I\approx \eta \sum_{m\in \Lambda_\textrm{B}'\setminus\{0\}} g_{\textrm{d}}(l_m) \xi_m.
\end{equation}
For simplicity, to derive the Laplace transform in the sequel,
\rev{we adopt the \textbf{Approximation 5} that $\eta$ is approximately replaced with its mean value for the three cases conditioned on $d_0$ (i.e., with IRS reflect beamforming, with IRS scattering only, and without nearby IRS), respectively,}\footnote{\rev{As discussed in Section \ref{SectionSignalWithout_d0}, the nearest IRS 0 (and hence the distance $d_0$) typically has the dominant impact on the randomly scattered signal or interference compared to other (farther) IRSs, whereas such impact also decays quickly as $d_0$ increases. Therefore, \textbf{Approximation 5} is practically reasonable, which is also verified by simulation results in Section \ref{SectionSimulation}.}} which is given by
\begin{equation}\label{bareta}
	\bar \eta\triangleq \begin{cases}
	\kappa_\textrm{sc}(d_0), & \textrm{if} \ d_0\leq D_1;\\
	1+N E_\textrm{I1}(D_1), & \textrm{if} \ D_1<d_0\leq D_2;\\
	1, & \textrm{otherwise.}
	\end{cases}
\end{equation}
Note that $\bar\eta$ as a function of $d_0$ is in general not continuous at the boundary point $D_1$ (or $D_2$), similar to the case in \eqref{Sgamma}.
As a result, we have the following proposition.
\begin{prop}\label{Prop_LaplaceI}
\rev{Based on \textbf{Approximations 1}, \textbf{2}, \textbf{4} and \textbf{5},} the Laplace transform of the conditional interference power $I|_{l_0,d_0}$ is given by
\begin{align}
&\mathcal{L}_{I|_{l_0,d_0}}(s)\triangleq \mathbb{E}\{e^{-sI}\}|_{l_0,d_0}\approx\exp\big(-2\pi\lambda_\textrm{B}' U(\bar \eta s) \big),\label{Ilp}
\end{align}
where the function $U(\cdot)$ is defined as
\begin{align}
U(x)\triangleq & \frac{\pi}{\alpha \sin(\frac{2\pi}{\alpha})}(\beta x)^{\frac{2}{\alpha}}\notag\\
&- \frac{l_0^2+H_\textrm{B}^2}{2}\cdot {}_{2}F_{1}\bigg(1,\frac{2}{\alpha},1+\frac{2}{\alpha},-\frac{1}{g_\textrm{d}(l_0) x}\bigg),\label{Ux}
\end{align}%
with $g_\textrm{d}(l_0)=g_\textrm{d}(l)|_{l=l_0}$ given by \eqref{gd}, and ${}_{2}F_{1}$ denoting the Gauss hypergeometric function\cite{NISTfunctions}.
\end{prop}
\textit{Proof:} Please refer to Appendix \ref{AppendixLaplace}.$\blacksquare$

Finally, the cdf of the conditional interference power $I|_{l_0,d_0}$ can be obtained by taking the inverse Laplace transform of \eqref{Ilp}, i.e.,
\begin{equation}\label{IcdfCond}
	F_{I|_{l_0,d_0}}(x)=\mathcal{L}^{-1}\bigg[\frac{1}{s}\mathcal{L}_{I|_{l_0,d_0}}(s)\bigg](x),
\end{equation}
which can be computed directly in MATLAB.

\subsection{Mean Interference Power}\label{SectionMeanI}

Based on \eqref{Iapprox}, \eqref{bareta} and i.i.d. $\xi_m\sim\xi\sim \textrm{Exp}(1), \forall m$, the mean interference power conditioned on $l_0$ and $d_0$ is given by 
\begin{align}\label{Imean}
&\mathbb{E}\{I\}|_{l_0,d_0}\approx\mathbb{E}\{\xi\} \mathbb{E}\{\eta\}|_{d_0}\mathbb{E}\bigg\{\sum_{m\in \Lambda_\textrm{B}'\setminus\{0\}} g_{\textrm{d}}(l_m)\bigg\}\bigg|_{l_0}\notag\\
&=\mathbb{E}\{\eta\}|_{d_0} E_\textrm{B1}(l_0),
\end{align}
where $E_\textrm{B1}(l_0)$ represents the expectation of the sum of direct channel powers from interfering BSs $m\in \Lambda_\textrm{B}'\setminus\{0\}$ conditioned on $l_0$, which is given by
\begin{align}
&E_\textrm{B1}(l_0)\triangleq \mathbb{E}\bigg\{\sum_{m\in \Lambda_\textrm{B}'\setminus\{0\}} g_{\textrm{d}}(l_m)\bigg\}\bigg|_{l_0}\stackrel{(b)}{=}2\pi\lambda_\textrm{B}'\int_{l=l_0}^\infty g_{\textrm{d}}(l)l\diff l\notag\\
&=\frac{2\pi\lambda_\textrm{B}'\beta}{(\alpha-2)(l_0^2+H_\textrm{B}^2)^{\frac{\alpha}{2}-1}}.
\end{align}
Note that $(b)$ is due to the HPPP-distributed BS locations, which is essential to simplify the infinite interference summation as a spatial integral over the 2D plane.

Furthermore, we can integrate over the distribution of $l_0$ to obtain the conditional mean interference
\begin{equation}\label{MeanId0}
\mathbb{E}\{I\}|_{d_0}\approx \mathbb{E}\{\eta\}|_{d_0} E_\textrm{B2},
\end{equation}
with $E_\textrm{B2}$ defined as
\begin{align}
&E_\textrm{B2}\triangleq \mathbb{E}\big\{E_\textrm{B1}(l_0)\big\}=\int_{l_0=0}^\infty E_\textrm{B1}(l_0) f_{l_0}(l_0)\diff l_0\notag\\
&=\frac{2\pi\lambda_\textrm{B}'\beta}{(\alpha-2)}  \lambda_\textrm{B}\pi H_\textrm{B}^{4-\alpha} e^{\lambda_\textrm{B}\pi H_\textrm{B}^2} E_{\frac{\alpha}{2}-1}(\lambda_\textrm{B}\pi H_\textrm{B}^2),
\end{align}
where $f_{l_0}(l_0)$ is the pdf of $l_0$, and $E_{\frac{\alpha}{2}-1}(\cdot)$ is the exponential integral function with parameter $\frac{\alpha}{2}-1$.

In the case with $d_0\leq D_2$, we have $\mathbb{E}\{\eta\}|_{d_0}=\kappa_\textrm{sc}(d_0)$ and hence $\mathbb{E}\{I\}|_{d_0}$ in \eqref{MeanId0} is proportional to the IRS-dependent scattering gain $\kappa_\textrm{sc}(d_0)=1+N g_{\textrm{r}}(d_0) +N E_\textrm{I1}(d_0)$ in \eqref{kappasc}.
Note that the scattering gain coefficient $G_\textrm{sc}\triangleq N$ is generally not large enough to compensate the pathloss of the IRS-UE link. As a result, $\kappa_\textrm{sc}(d_0)$ is dominated by the preceding term of 1 that accounts for the direct BS-UE link, and the IRS-reflected interference is non-negligible only when $d_0$ is sufficiently small.
Moreover, similar to the analysis in Section \ref{SectionSignalWithout_d0}, the nearest IRS 0 (i.e., with the smallest distance $d_0$) has the dominant impact on the mean interference power compared to other IRSs.

Finally, based on \eqref{MeanId0} and the definition of $\eta$, the unconditional mean interference is given by
\begin{align}\label{ImeanUncond}
\mathbb{E}\{I\}\approx&\mathbb{E}\{\eta\}\mathbb{E}\bigg\{\sum_{m\in \Lambda_\textrm{B}'\setminus\{0\}} g_{\textrm{d}}(l_m)\bigg\}=\big(1+N E_\textrm{I1}(0)\big) E_\textrm{B2},
\end{align}
where $N\cdot E_\textrm{I1}(0)=N\cdot E_\textrm{I1}(d_0)\big|_{d_0=0}$ based on \eqref{EI1}, which represents the expectation of the sum of IRS-UE channel powers from all IRSs $j$ within horizontal distance $D_2$ from UE 0.


\section{Spatial Throughput Characterization}\label{SectionCharacterization}
In this section, we first derive the conditional SINR distribution and non-outage probability based on the conditional distributions of the signal power $S$ and interference power $I$ obtained in Sections \ref{SectionSignal} and \ref{SectionInterference}, respectively.
Then the coverage probability and hence spatial throughput are further obtained by integrating over the distributions of the distances $l_0$ and $d_0$.

In particular, the conditional non-outage probability is expressed in terms of the conditional interference power Laplace transform and its derivatives.
Note that although we can also approximate the conditional interference power distribution by the Gamma distribution and obtain closed-form expressions for the conditional non-outage probability based on the approach in \cite{RobertHeathGamma}, it is found by simulations that the resulted SINR distribution is not accurate in our considered setup and thus is not adequate to be used to further obtain the coverage probability that requires integration over the distributions of $l_0$ and $d_0$.

\subsection{Conditional SINR Distribution and Non-Outage Probability}\label{SectionSINR}

The non-outage probability conditioned on $l_0$ and $d_0$ is defined as
\begin{equation}\label{PnoCond}
\textrm{P}_{\textrm{no}}|_{l_0,d_0}\triangleq \mathbb{P}\{\gamma>\bar \gamma\}|_{l_0,d_0}=\mathbb{P}\{S>\bar \gamma(I +W)\}|_{l_0,d_0},
\end{equation}
which is related to the conditional SINR distribution via $\textrm{P}_{\textrm{no}}|_{l_0,d_0}=1-F_{\gamma}(\bar\gamma)|_{l_0,d_0}$, with $F_{\gamma}(\cdot)|_{l_0,d_0}$ denoting the conditional SINR cdf.

In Section \ref{SectionSignal}, we have approximated the conditional signal power distribution by the Gamma distribution $\Gamma[k_S,\theta_S]$ given in \eqref{Sgamma}.
With similar derivations in \cite{AndrewsGeneralFading}, we have the following lemma.
\newtheorem{lem}{Lemma}
\begin{lem}\label{lem_Pno}
	For a Gamma-distributed RV $S\sim \Gamma[k_S,\theta_S]$ with integer $k_S$ and an independent RV $X$, we have
	\begin{small}
	\begin{align}
	&\mathbb{P}\{S>X\}=\mathbb{E}_{X}\bigg\{\frac{\Gamma\big(k_S,\frac{X}{\theta_S}\big)}{\Gamma(k_S)}\bigg\}=\sum_{i=0}^{k_S-1} \frac{(-1)^i}{i!}\frac{\partial^i}{\partial s^i}\big[\mathcal{L}_{Y}(s)\big]_{s=1},
	\end{align}
	\end{small}
	where $\Gamma(\cdot,\cdot)$ is the upper incomplete Gamma function\cite{NISTfunctions}, and $Y=X/\theta_S$.
\end{lem}
\textit{Proof:} Please refer to Appendix \ref{AppendixPno}.$\blacksquare$

As a result, for integer $k_S$, the conditional non-outage probability in 
\eqref{PnoCond} is given by
\begin{align}\label{Pno}
&\textrm{P}_{\textrm{no}}|_{l_0,d_0}\approx\mathbb{E}_{I}\bigg\{\frac{\Gamma\big(k_S,\frac{\bar \gamma(I +W)}{\theta_S}\big)}{\Gamma(k_S)}\bigg\}\bigg|_{l_0,d_0}\notag\\
&=\sum_{i=0}^{k_S-1} \frac{(-1)^i}{i!}\frac{\partial^i}{\partial s^i}\big[\mathcal{L}_{Y|_{l_0,d_0}}(s)\big]_{s=1},
\end{align}
where $Y\triangleq \frac{\bar \gamma(I +W)}{\theta_S}$.
Based on \eqref{Ilp}, the Laplace transform of $Y|_{l_0,d_0}$ is given by
\begin{align}
&\mathcal{L}_{Y|_{l_0,d_0}}(s)\triangleq \mathbb{E}\big\{e^{-sY}\big\}\big|_{l_0,d_0}\notag\\
&=\exp\bigg(-\frac{s\bar \gamma W }{\theta_S}\bigg) \mathcal{L}_{I|_{l_0,d_0}}\bigg(\frac{s\bar \gamma }{\theta_S}\bigg)=\exp\big(V(s)\big),
\end{align}
where $V(s)\triangleq -\frac{s\bar \gamma W }{\theta_S} -2\pi\lambda_\textrm{B}' U\big( \frac{s\bar \gamma \bar \eta}{\theta_S}\big)$.
In order to evaluate $\textrm{P}_{\textrm{no}}|_{l_0,d_0}$ in \eqref{Pno}, the first $(k_S-1)$-order derivatives of the composite function $\exp(V(s))$ are needed, which are derived in Appendix \ref{AppendixVs}.

In the above, we have addressed the case with integer $k_S$. For non-integer $k_S$, we can obtain the conditional non-outage probability for the cases with the upper and lower integers of $k_S$, respectively, and approximate $\textrm{P}_{\textrm{no}}|_{l_0,d_0}$ by their linear interpolation, i.e.,
\begin{equation}
	\textrm{P}_{\textrm{no}}|_{l_0,d_0}\approx w\textrm{P}_{\textrm{no}}|_{l_0,d_0,\lfloor k_S\rfloor}+ (1-w) \textrm{P}_{\textrm{no}}|_{l_0,d_0,\lceil k_S\rceil},
\end{equation}
where $\lfloor \cdot\rfloor$ and $\lceil \cdot\rceil$ denote the floor and ceiling functions, respectively, and the weight $w$ is given by
\begin{equation}
w\triangleq \frac{M(\lceil k_S\rceil -k_S)}{M(\lceil k_S\rceil -k_S) +(k_S-\lfloor k_S\rfloor )}.
\end{equation}
Note that the weights $w$ and $(1-w)$ are designed such that their ratio is proportional to the ratio of the distances from $k_S$ to its upper and lower integers, respectively, i.e., $\frac{w}{1-w}=M\frac{\lceil k_S\rceil -k_S}{k_S-\lfloor k_S\rfloor}$, with $M>0$ denoting the priority factor to account for the nonlinearity of $\textrm{P}_{\textrm{no}}|_{l_0,d_0}$ with $k_S$.

Finally, for the case with reflect beamforming, the value of $k_S$ increases as $d_0$ decreases (due to channel hardening effect discussed in Section \ref{SectionSignalMean}), which requires more higher order derivatives of the Laplace transform in \eqref{Pno} to attain good accuracy.
Fortunately, we notice the fact that for a large value of $k_S$, the Gamma distribution approaches the normal distribution whose mean value $\mu\triangleq\mathbb{E}\{S\}|_{l_0,d_0}$ is much larger than its standard deviation $\omega\triangleq \sqrt{\var\{S\}|_{l_0,d_0}}$, and hence we can approximately represent the signal power by its mean value $\mu$.
Therefore, when $k_S$ is larger than a certain threshold $\tilde{k}_S$, we have
\begin{equation}
\textrm{P}_{\textrm{no}}|_{l_0,d_0}=\mathbb{P}\{I< S/\bar \gamma -W\}|_{l_0,d_0}\approx F_{I|_{l_0,d_0}}(z),
\end{equation}
where $z\triangleq \mu/\bar \gamma -W$ and $F_{I|_{l_0,d_0}}(\cdot)$ is the conditional cdf of $I$ given by \eqref{IcdfCond}.

\subsection{Coverage Probability and Spatial Throughput}\label{SectionErgodic}

After obtaining the conditional non-outage probability $\textrm{P}_{\textrm{no}}|_{l_0,d_0}$, we can then obtain the coverage probability, i.e., average non-outage probability in the network, by integrating over the distributions of $l_0$ and $d_0$, given by
\begin{align}\label{IntegralPno}
\textrm{P}_{\textrm{cov}}=& \int_{l_0=0}^{\infty}\int_{d_0=0}^\infty \textrm{P}_{\textrm{no}}|_{l_0,d_0} f_{d_0}(d_0) f_{l_0}(l_0) \diff d_0\diff l_0\notag\\
=&\int_{l_0=0}^{\infty}\int_{d_0=0}^{D_1} \textrm{P}_{\textrm{no}}|_{l_0,d_0} f_{d_0}(d_0)f_{l_0}(l_0) \diff d_0 \diff l_0 \notag\\
&+\textrm{P}_{\textrm{sc}}\int_{l_0=0}^{\infty} \textrm{P}_{\textrm{no}}|_{l_0,D_1<d_0\leq D_2} f_{l_0}(l_0)\diff l_0\notag\\
& +\textrm{P}_{\textrm{wo}}\int_{l_0=0}^{\infty} \textrm{P}_{\textrm{no}}|_{l_0,d_0> D_2} f_{l_0}(l_0)\diff l_0,
\end{align}%
where $\textrm{P}_{\textrm{no}}|_{l_0,D_1<d_0\leq D_2}$ and $\textrm{P}_{\textrm{no}}|_{l_0,d_0> D_2}$ denote the conditional non-outage probabilities for the cases with $D_1<d_0\leq D_2$ and $d_0> D_2$, which occur with probabilities $\textrm{P}_{\textrm{sc}}$ and $\textrm{P}_{\textrm{wo}}$ given in Section \ref{SectionSignalMean}, respectively.
The integration is divided into three parts based on $d_0$, i.e., with IRS reflect beamforming, with IRS scattering only, and without nearby IRS.
Note that the integrands in \eqref{IntegralPno} admit closed forms and thus the integral can be evaluated efficiently.
Finally, the spatial throughput in \eqref{SpatialThroughput} can be obtained as $\nu=\bar R p\lambda_\textrm{B} \textrm{P}_{\textrm{cov}}$.

\section{Numerical Results}\label{SectionSimulation}

In this section, we verify our analytical results by Monte-Carlo (MC) simulations for the conditional signal/interference power distribution in \eqref{Sgamma}/\eqref{IcdfCond}, the mean signal/interference power in \eqref{Sint}/\eqref{ImeanUncond}, the coverage probability in \eqref{IntegralPno} and the spatial throughput in \eqref{SpatialThroughput}, and investigate the impact of key system parameters including the BS/IRS densities, IRS element number $N$ and network loading factor $p$.
Each MC simulation result is obtained by averaging over 2000 randomly generated topologies in a disk area of radius 20 kilometers (km), with 1000 fading channel realizations per channel. Based on the results presented in this section, it is verified that our analytical results match well with the MC simulation results.
Denote $\lambda_0=5\times 10^{-6}$ /m$^2$ as the BS or IRS reference density.\footnote{For example, $\lambda_\textrm{B}=\lambda_0$ corresponds to an average cell radius of 252 m if each cell is approximated by a disk region.}
The following parameters are used if not mentioned otherwise: 
$H_\textrm{B}=20$ m, $H_\textrm{I}=1$ m,\footnote{The parameter $H_\textrm{I}=1$ m corresponds to the case where the IRS is deployed at a relative height of 1 meter above the UE level (i.e., 0 meter of altitude), which can also be chosen as any practical value of interest.} $W=-147$ dB, $f_c=2$ GHz, $\alpha=3$, $\tilde{k}_S=8$, $M=1.5$, $\bar R=1$ bps/Hz, $D_1=25$ m and $D_2=50$ m.

\subsection{Performance with Given BS Density}

\subsubsection{Conditional Signal/Interference Power Distribution}\label{SectionSimulationConditional}
\begin{figure}
	\centering
	\includegraphics[width=1\linewidth,  trim=40 0 60 10,clip]{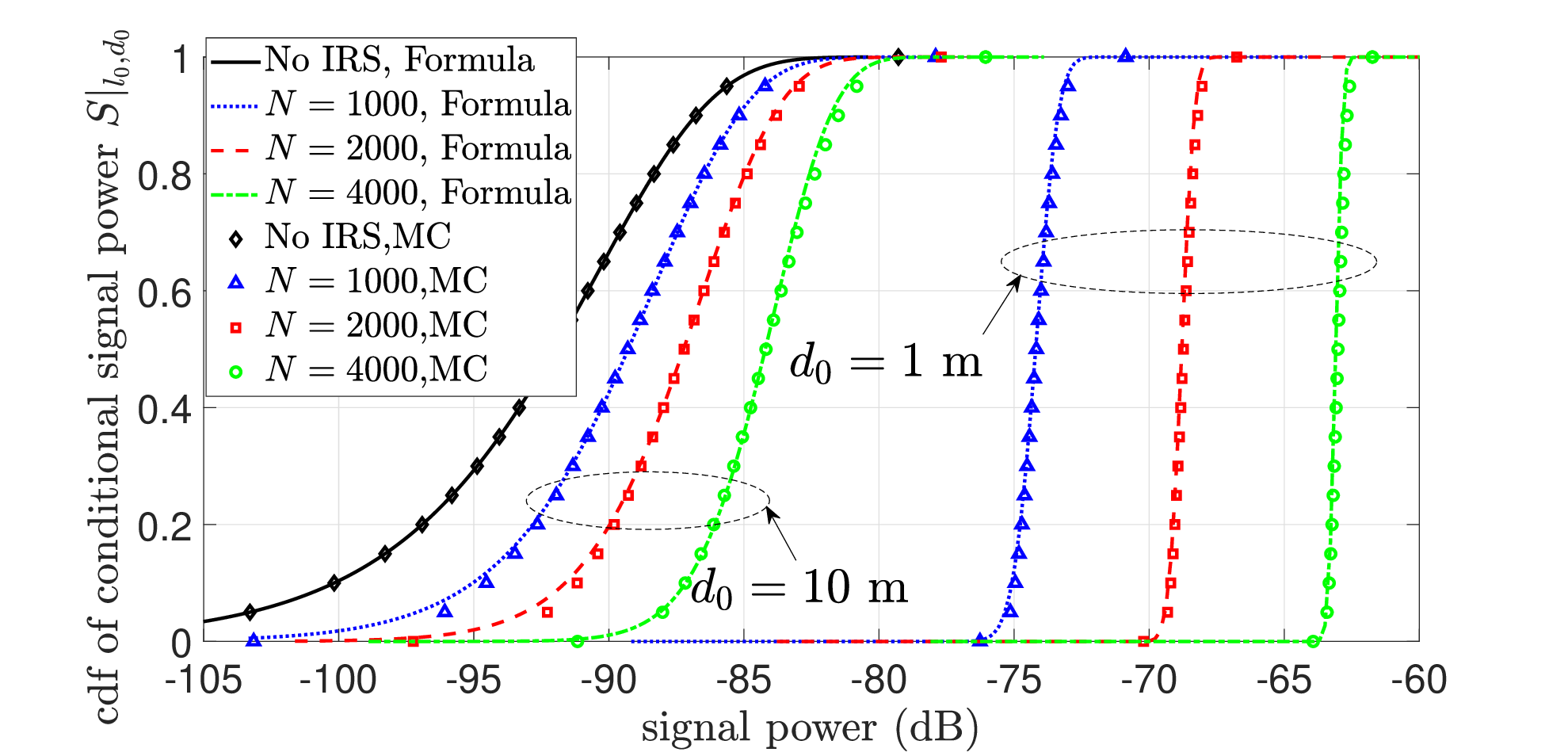}
	\caption{Cdf of the conditional signal power $S|_{l_0,d_0}$ with reflect beamforming from the associated IRS, under $\lambda_\textrm{B}=10\lambda_0$, $\lambda_\textrm{I}=100\lambda_0$, $l_0=50$ m and different $d_0$ and $N$.\vspace{-2ex}}\label{Scdf}
\end{figure}

\begin{figure}
	\centering
	\includegraphics[width=1\linewidth,  trim=60 0 60 8,clip]{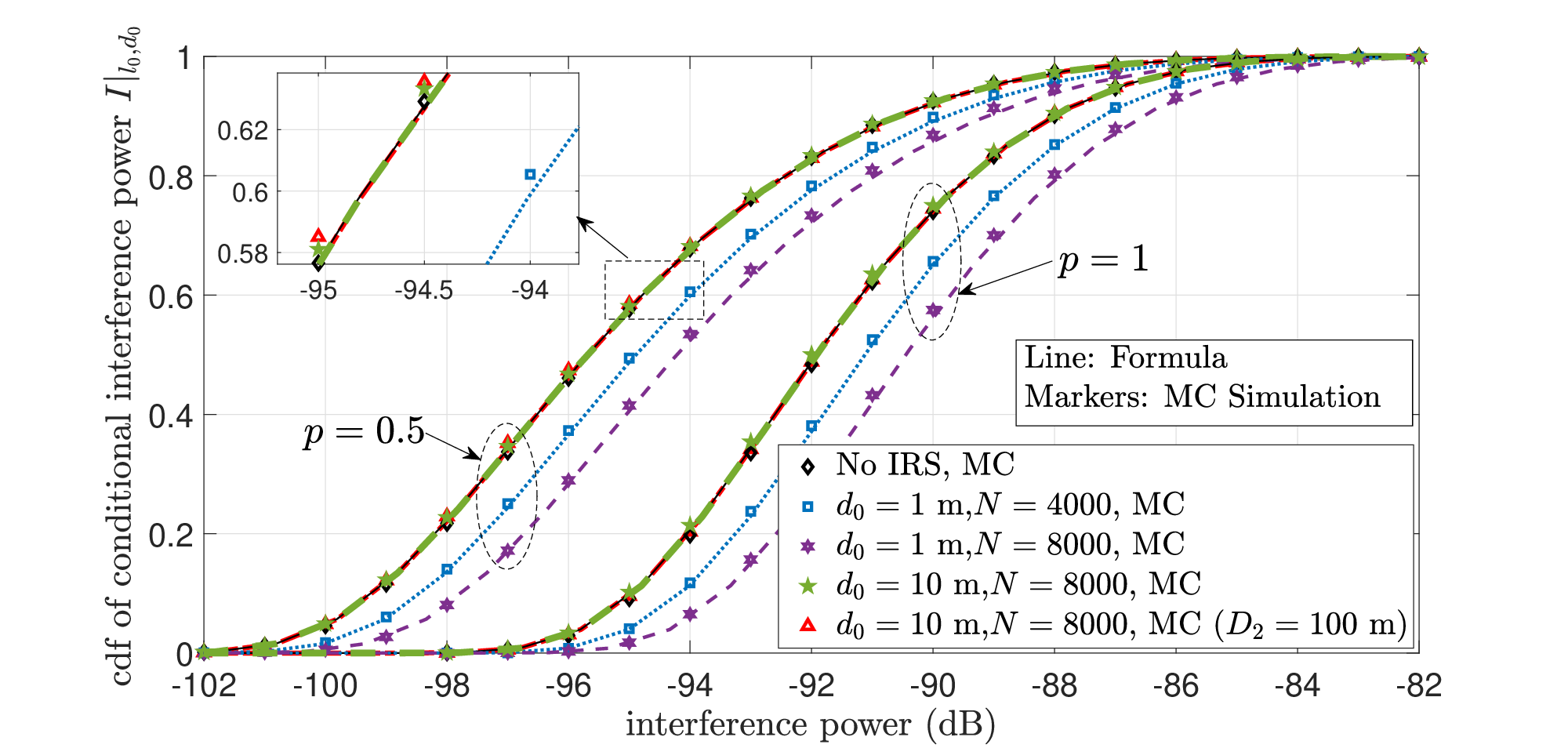}
	\caption{\rev{Cdf of the conditional interference power $I|_{l_0,d_0}$, under $\lambda_\textrm{B}=10\lambda_0$, $\lambda_\textrm{I}=100\lambda_0$, $l_0=50$ m and different $d_0$, $N$ and $p$.}\vspace{-2ex}}\label{Icdf}
\end{figure}

The cdf of the conditional signal/interference power in \eqref{Sgamma}/\eqref{IcdfCond} under $\lambda_\textrm{B}=10\lambda_0$, $\lambda_\textrm{I}=100\lambda_0$, $l_0=50$ m and different $d_0$ and $N$ is plotted in Fig. \ref{Scdf} and Fig. \ref{Icdf}, respectively.
It is observed from Fig. \ref{Scdf} that, with reflect beamforming from the associated IRS 0, the signal power is significantly enhanced and exhibits channel hardening when $N$ increases and/or the IRS 0-UE 0 horizontal distance $d_0$ decreases. 

\rev{For the interference power distribution shown in Fig. \ref{Icdf}, the IRSs within range $D_2$ of the target UE 0 randomly scatter the signal from interfering BSs, which only slightly increase the interference power even when $d_0$ is very small (e.g., $d_0=1$ m) and $N$ is very large (e.g., $N=8000$), compared to the case without IRS.
Moreover, as $d_0$ increases (e.g., $d_0\geq 10$ m), it can be seen that there is negligible difference on the conditional interference power distribution for cases without versus with IRS (even when $N=8000$ and under larger $D_2$, e.g., $D_2=100$ m).
These observations validate our analysis in Section \ref{SectionSignalWithout_d0} that under practical values of the IRS density, the nearest IRS 0 (and hence the distance $d_0$) typically has the dominant impact on the randomly scattered signal or interference compared to other (farther) IRSs, whereas such impact also decays quickly as $d_0$ increases.
This could be attributed to the severe product-distance/double-pathloss attenuation of the signal/interference randomly reflected by IRS.
}
In contrast, the interference power significantly increases when the network loading factor $p$ increases (corresponding to more UEs per RB) regardless of with or without IRSs in the network, since more co-channel/interfering BSs become active on the same RB.

\subsubsection{Impact of $\lambda_\textrm{I}$ and $N$ on Coverage Probability}

\begin{figure}
	\centering
	\includegraphics[width=1\linewidth,  trim=40 0 60 0,clip]{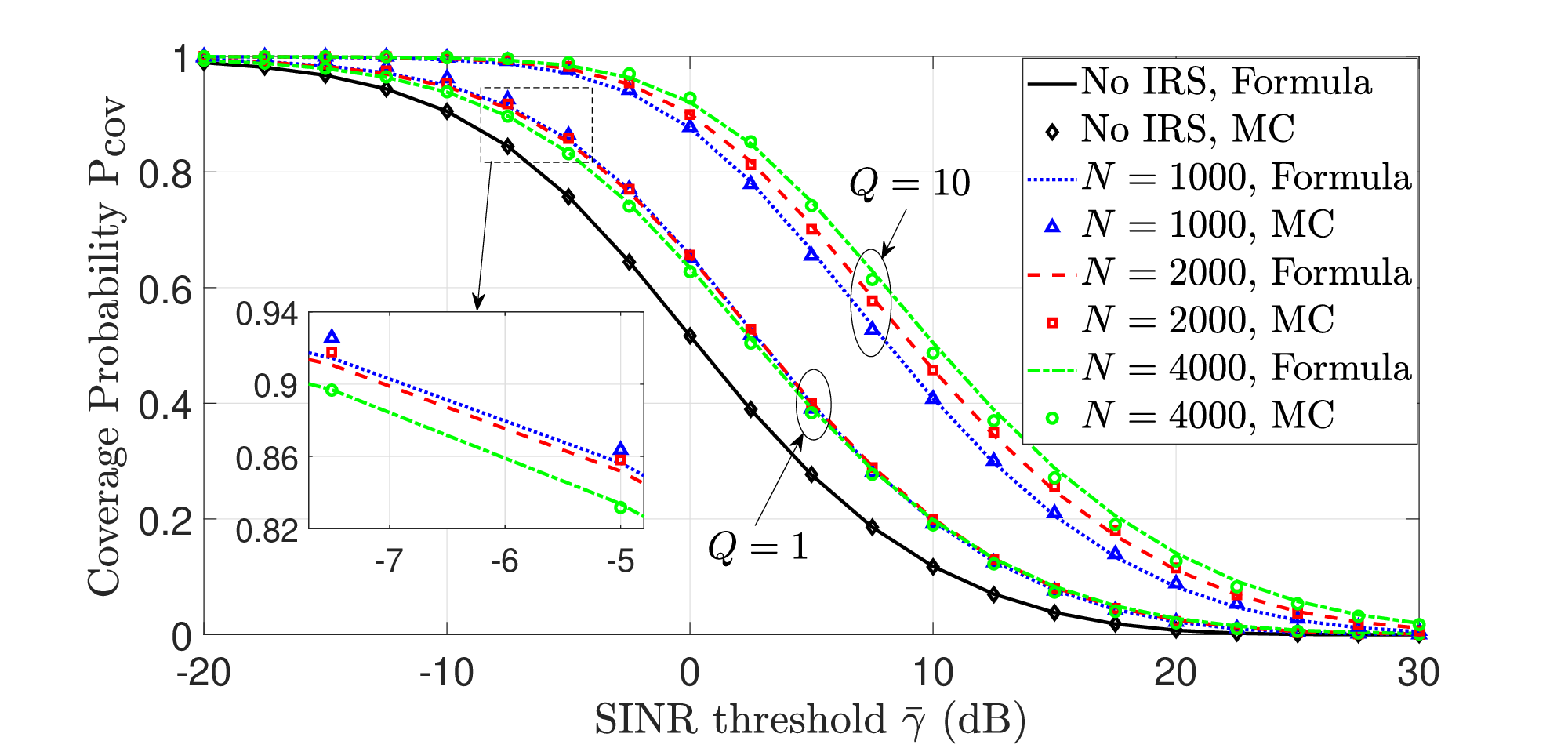}
	\caption{Coverage probability with different $Q$ and $N$, under $\lambda_\textrm{B}=10\lambda_0$ and $p=0.5$.\vspace{-2ex}}\label{PcovSINR}
\end{figure}

Next, we investigate the impact of the IRS density $\lambda_\textrm{I}$ and element number $N$ on the network coverage probability $\textrm{P}_{\textrm{cov}}$ given by \eqref{IntegralPno}.
Specifically, denote $Q\triangleq N \lambda_\textrm{I}$ as the total number of IRS elements per m$^2$.
The coverage probability under different $Q$ and $N$ is plotted in Fig. \ref{PcovSINR} for comparison. 

First, it is observed that increasing $Q$ helps improve the coverage probability.
Second, in the case with small $Q$ (e.g., $Q=1$/m$^2$), choosing smaller $N$ under the same $Q$ leads to slightly higher coverage probability at the low-SINR threshold region (see the zoomed plot inside Fig. \ref{PcovSINR}). 
The reason is that, under the same $Q$, a smaller $N$ leads to higher IRS density $\lambda_\textrm{I}$, which helps cover more UEs with low-SINR requirement.
In contrast, in the case with large $Q$ (e.g., $Q=10$/m$^2$), choosing higher $N$ under the same $Q$ yields higher coverage probability in both low and high SINR threshold regions.
This is because in this case $\lambda_\textrm{I}$ is already sufficiently large to cover most UEs and thus the IRS passive beamforming gain that grows with $N$ in $O(N^2)$ is more effective in enhancing UEs' signal power and hence the coverage probability.

\subsubsection{Mean Signal/Interference Power and Spatial Throughput}

\begin{figure}
	\centering
	\includegraphics[width=1\linewidth, trim=40 0 60 10,clip]{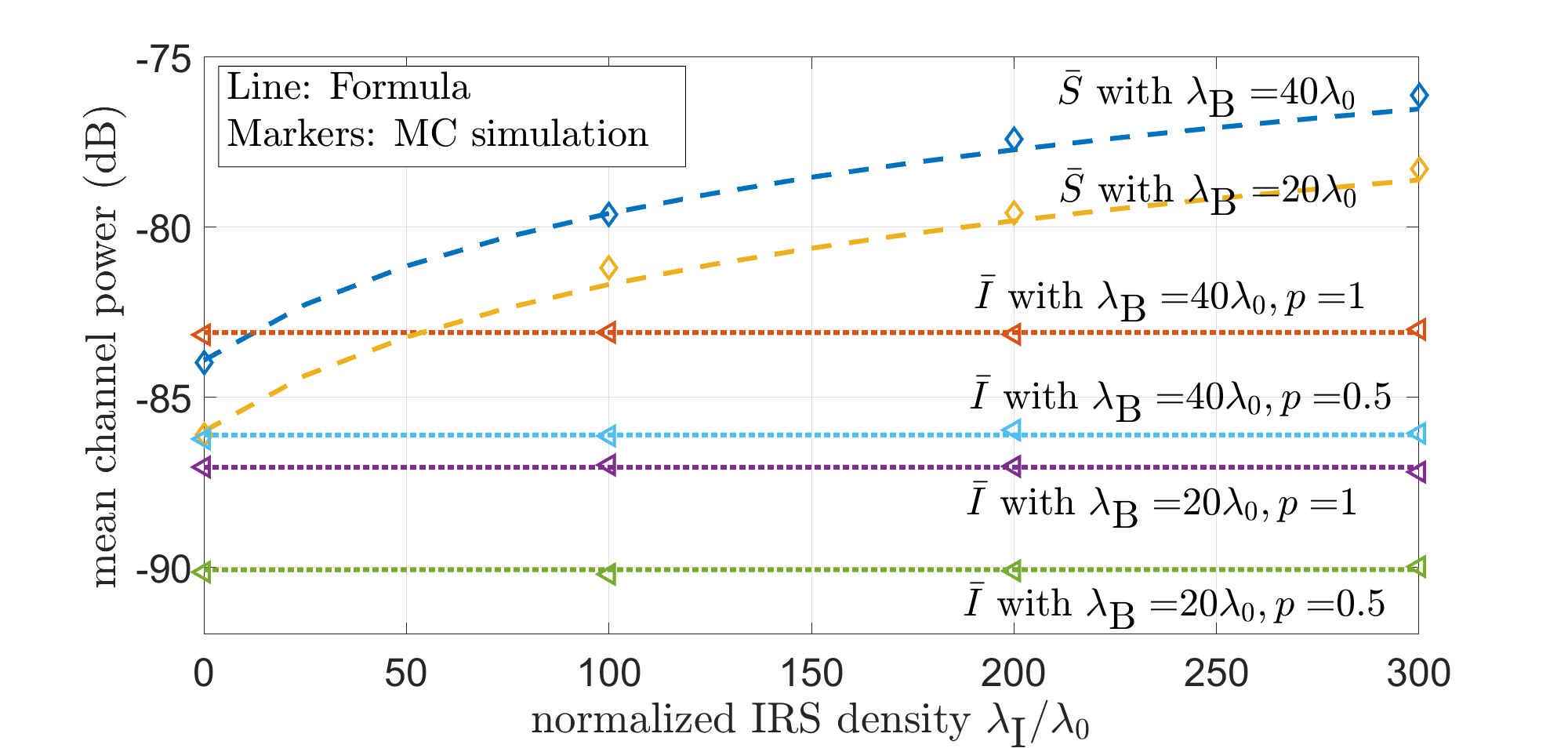}
	\caption{Mean signal (interference) power under different BS/IRS densities and loading factor $p$, with $N=2000$.\vspace{-2ex}}\label{SI_lambdaI}
\end{figure}

The mean signal/interference power in \eqref{Sint}/\eqref{ImeanUncond} under different BS/IRS densities and loading factor $p$ is plotted in Fig. \ref{SI_lambdaI} for comparison.
First, it is observed that for the case with BS only (i.e., $\lambda_\textrm{I}=0$), increasing $\lambda_\textrm{B}$ from $20\lambda_0$ to $40\lambda_0$ brings 2.1 dB (4.0 dB) gain to the mean signal (interference) power. In other words, when the BS density is large, the mean interference power increases faster than the mean signal power by adding more BSs.
In contrast, for the hybrid BS/IRS network, as the IRS density increases, the mean signal power increases significantly while the mean interference power increases only marginally. This is mainly due to the different scaling laws of the mean BS-IRS-UE channel power for the cases with and without IRS reflect beamforming as discussed in Section \ref{Statistics}, which are also consistent with the earlier observations made on Figs. \ref{Scdf} and \ref{Icdf}.

\begin{figure}
	\centering
	\includegraphics[width=1\linewidth,  trim=30 0 60 8,clip]{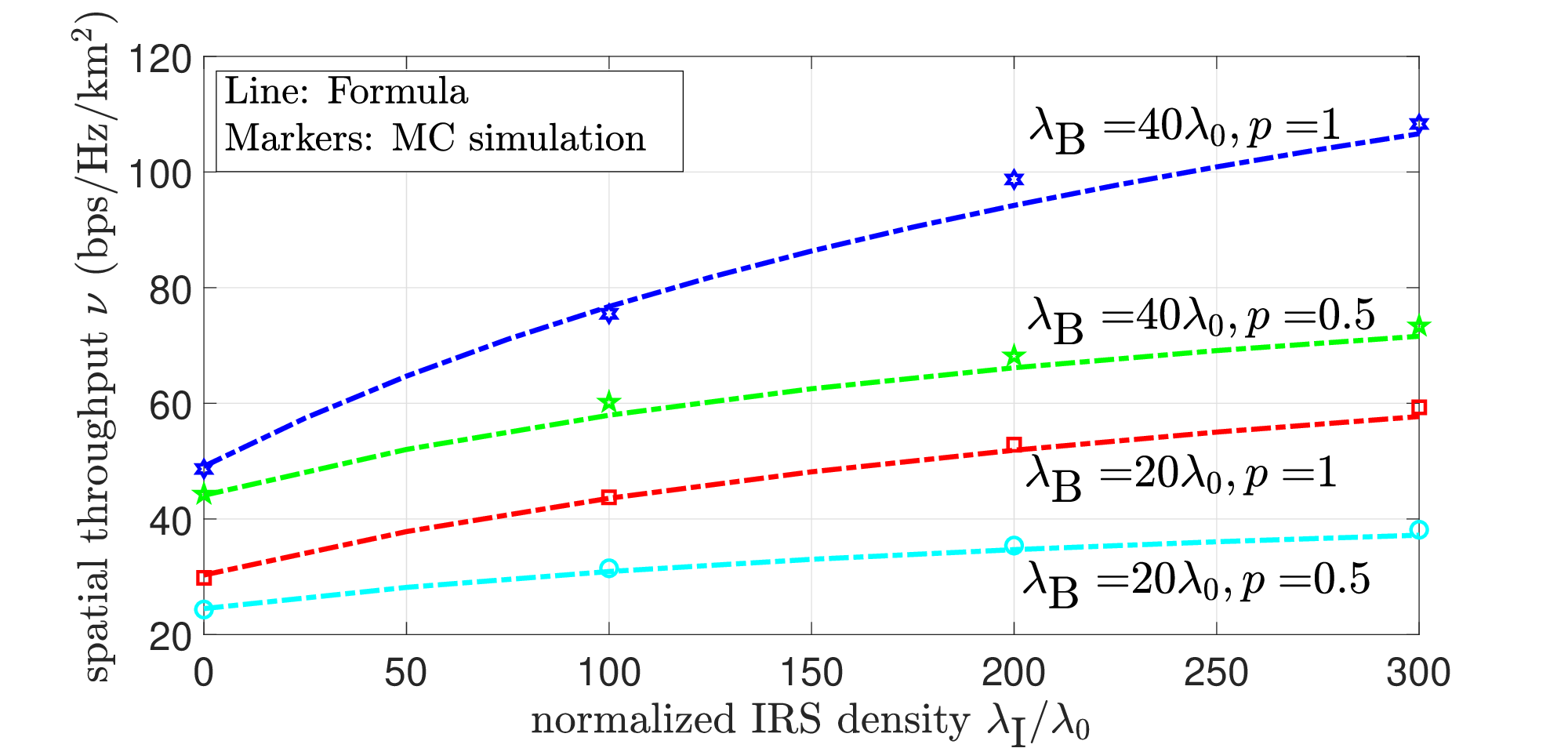}
	\caption{Spatial throughput $\nu$ under different BS/IRS densities and loading factor $p$, with $N=2000$.\vspace{-2ex}}\label{nu_lambdaI}
\end{figure}

Next, the spatial throughput $\nu$ under different BS/IRS densities and loading factor $p$ is plotted in Fig. \ref{nu_lambdaI}.
It is observed that, under given BS density, increasing IRS density always enhances the spatial throughput. 
Moreover, the speed of such increase is faster under a higher BS density $\lambda_\textrm{B}$ and/or higher network loading factor $p$.
The reasons are two-fold.
First, increasing IRS density helps enhance the signal power with only marginally increased interference power, as shown in Fig. \ref{SI_lambdaI}.
Second, a higher BS density shortens the BS-IRS distance on average and thus helps enhance the IRS reflected signal power for its served UEs, while a higher network loading factor implies more UEs per RB and thus more UEs served by each IRS on average, both leading to more substantial spatial throughput improvement with increasing IRS density.

\subsection{Spatial Throughput Subject to Total BS/IRS Cost}

In this subsection, we investigate the network spatial throughput subject to a given total cost of BSs and IRSs deployed.
Denote $c_0$ as the cost of each BS, and assume that the cost of each IRS with $N$ elements is $c_{\textrm{IRS},N}\triangleq c_0/K_N$, where $K_N>0$ is the BS/IRS cost ratio which could be any positive value according to practical BS and IRS costs. Nevertheless, the cost of one IRS is expected to be lower than one BS, thanks to the IRS's passive signal reflection without the need of any signal processing/regeneration\cite{QQirsMag}\cite{IRSbasar}. 
Denote $\zeta\triangleq \lambda_\textrm{I}/\lambda_\textrm{B}$ as the IRS/BS density ratio.
The total cost per m$^2$ in the IRS-aided hybrid wireless network is then given by
\begin{align}
C&\triangleq \lambda_\textrm{B} c_0 + \lambda_\textrm{I} c_{\textrm{IRS},N}=\lambda_\textrm{B} c_0 + \lambda_\textrm{I} c_0/K_N \notag\\
&=\lambda_\textrm{B} c_0 + \zeta\lambda_\textrm{B} c_0/K_N = \lambda_\textrm{B} c_0(1+\zeta/K_N).
\end{align}%
\vspace{-1.5em}

\begin{figure}
	\centering
	\includegraphics[width=1\linewidth,  trim=10 0 40 0,clip]{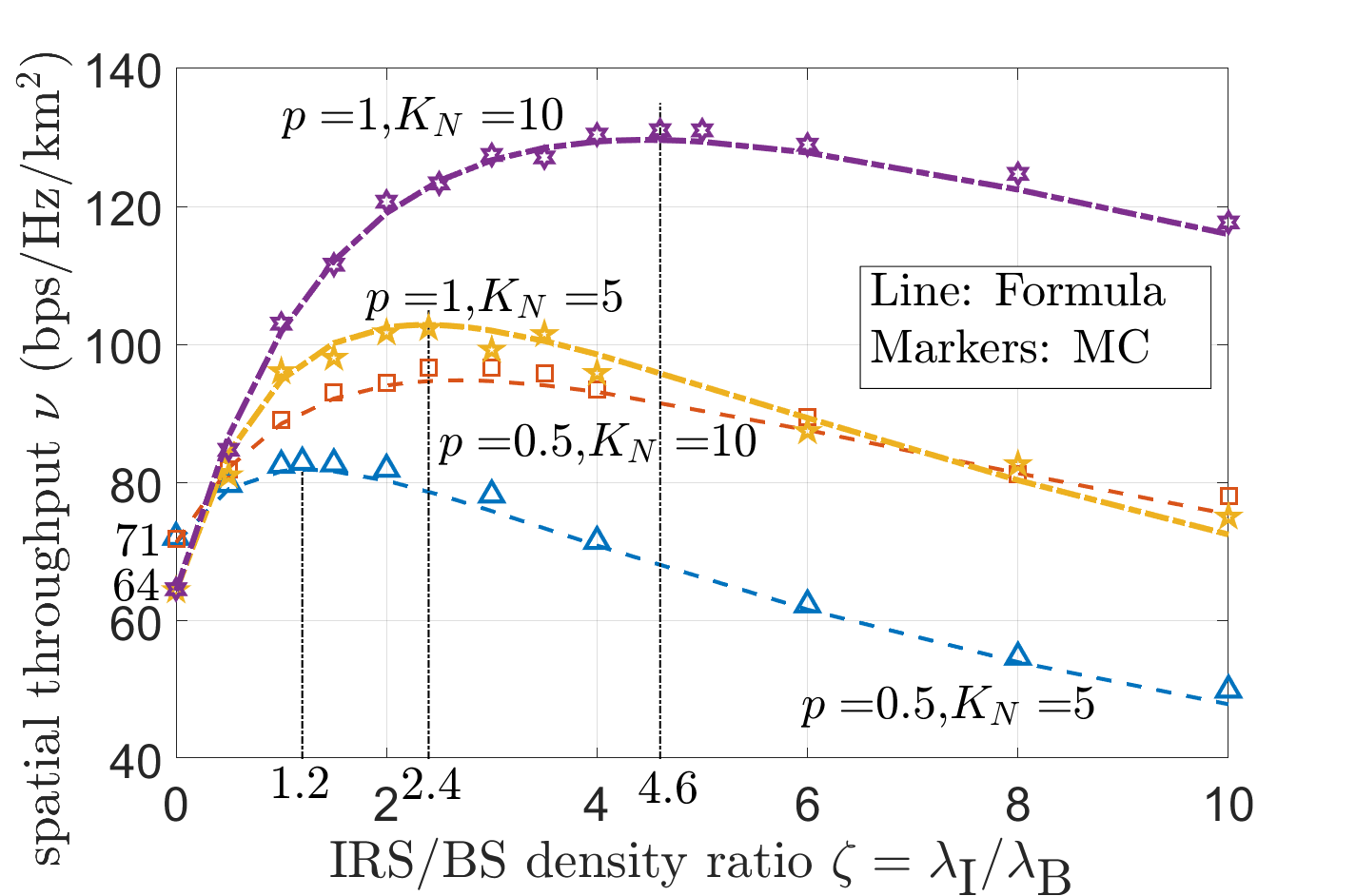}
	\caption{Spatial throughput $\nu$ versus IRS/BS density ratio $\zeta$ with total cost $C=80 \lambda_0 c_0$, under different BS/IRS cost ratio $K_N$ (with $N=2000$) and network loading factor $p$.\vspace{-2ex}}\label{nuRatio}
\end{figure}

The spatial throughput $\nu$ versus the IRS/BS density ratio $\zeta$ under given total cost $C$ is shown in Fig. \ref{nuRatio}.
First, it is observed that given the total cost $C$, there exists an optimal IRS/BS density ratio $\zeta^*$ that attains the maximum spatial throughput $\nu^*$, which is significantly higher than that of the BS-only network (i.e., $\zeta=0$) as well as the hybrid network with excessively large $\zeta$ (e.g., $\zeta=10$), where the BS density is too low to provide enough signal power for effective IRS reflection and passive beamforming. 
Second, it is observed that the optimal ratio $\zeta^*$ is roughly proportional to the BS/IRS cost ratio $K_N$ and the network loading factor $p$, where a larger $K_N$ suggests that more IRSs should be deployed each with relatively lower cost as compared to BS, while a higher $p$ corresponds to higher traffic demand (or more UEs per RB) which thus requires deploying more IRSs.

\begin{figure}
	\centering
	\includegraphics[width=1\linewidth,  trim=20 0 0 10,clip]{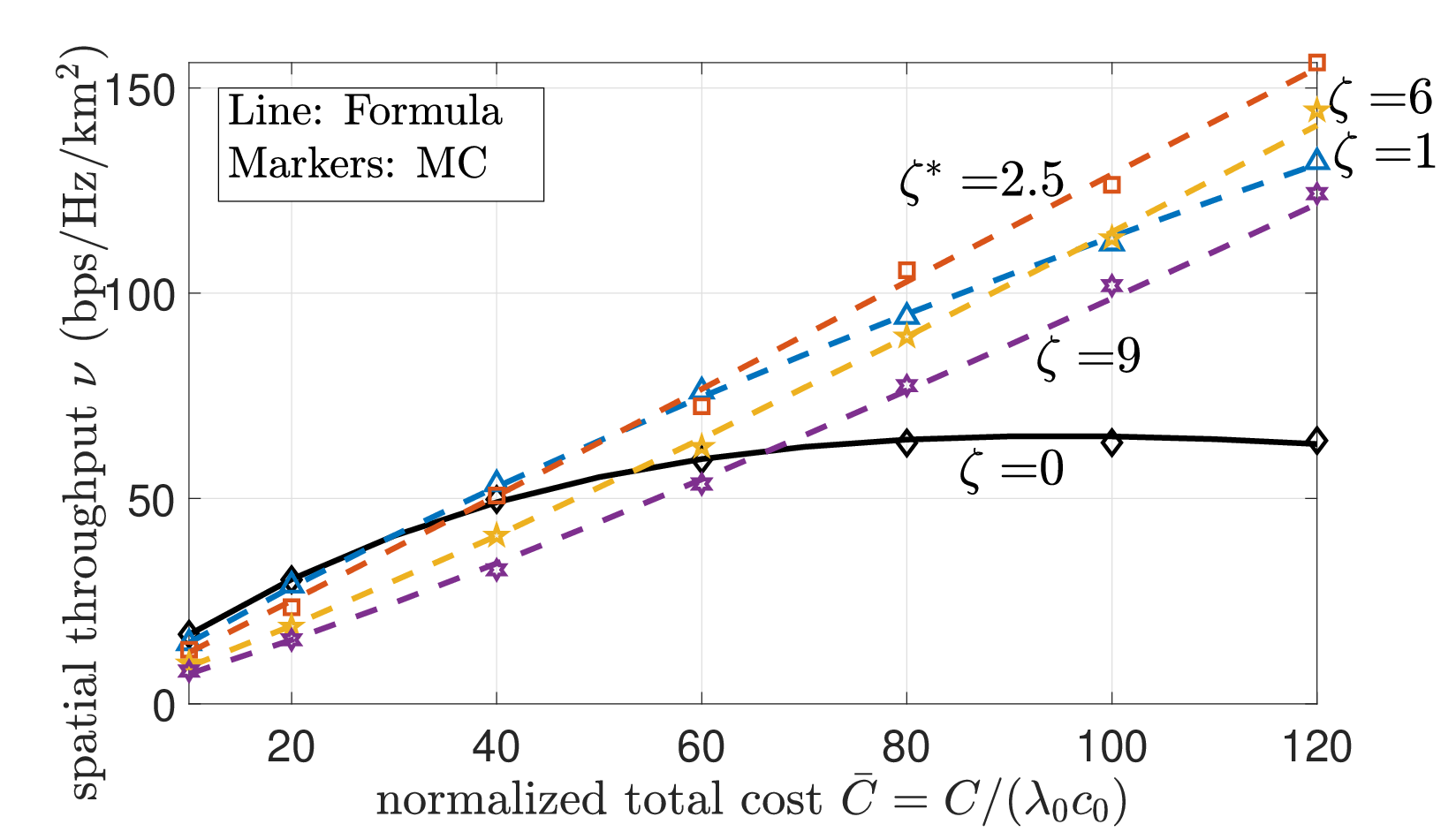}
	\caption{Spatial throughput $\nu$ versus total cost $C$ under different IRS/BS density ratio $\zeta$, with $p=1$ and given $K_N=5$ (with $N=2000$).\vspace{-2ex}}\label{nuC}
\end{figure}

Next, the spatial throughput $\nu$ versus total cost $C$ under different IRS/BS density ratio $\zeta$ is plotted in Fig. \ref{nuC}, for the case with $p=1$ and $K_N=5$.
First, it is observed that for the BS-only network (i.e., $\zeta=0$), the spatial throughput first increases and then decreases as the BS density increases, due to the more severe interference as compared to the improved signal power.
Second, when the BS density is larger than a certain value (say, $40\lambda_0$), the optimal IRS/BS density ratio $\zeta^*$ is approximately 2.5, while the maximum $\nu^*$ increases almost linearly with the total cost $C$, which significantly outperforms the BS-only network.

\begin{figure}
	\centering
	\includegraphics[width=1\linewidth,  trim=10 0 10 0,clip]{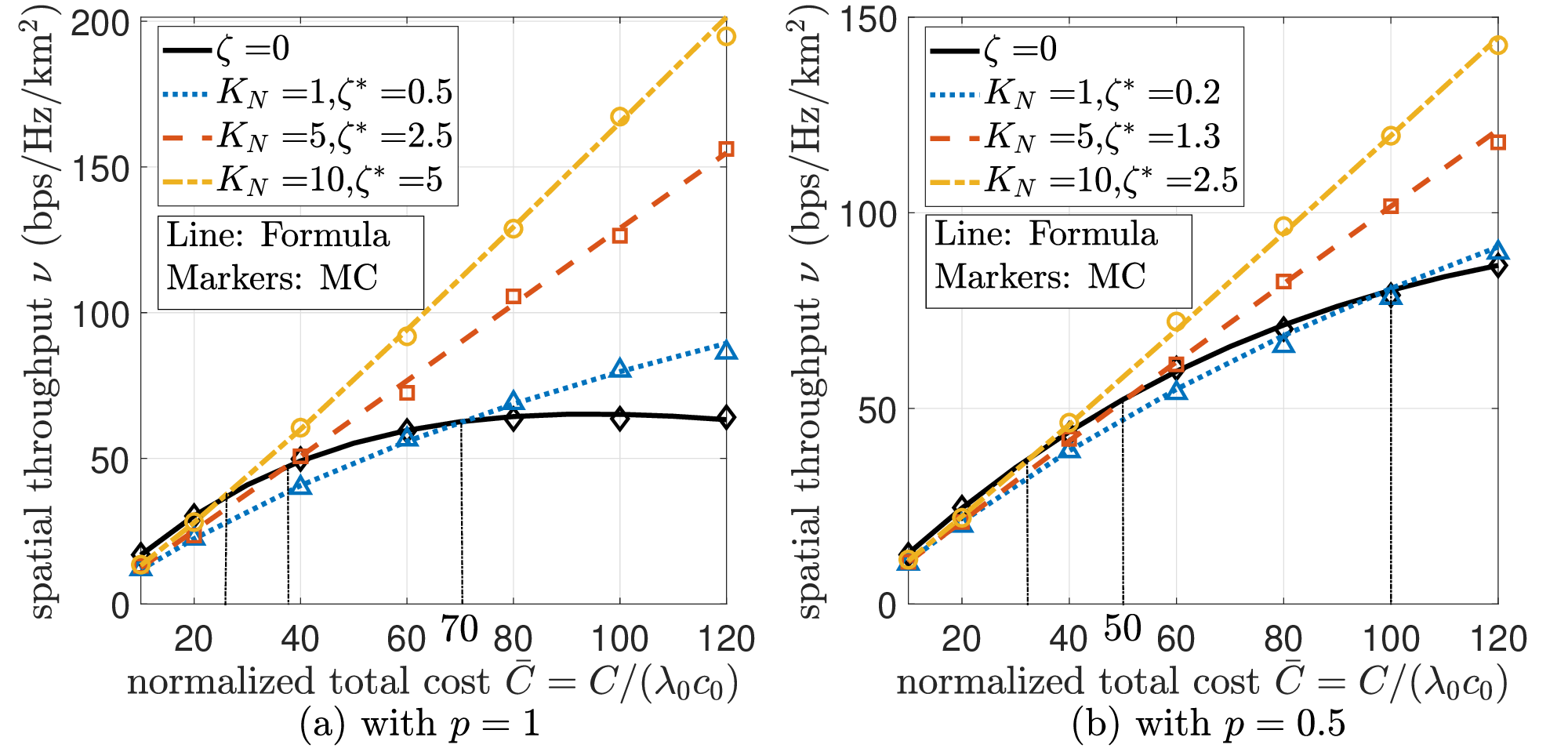}
	\caption{Spatial throughput $\nu$ versus total cost $C$ for (a) $p=1$ and (b) $p=0.5$, under different $K_N$ (with $N=2000$) and the corresponding optimal IRS/BS density ratio $\zeta^*$.\vspace{-2ex}}\label{nuCopt}
\end{figure}

Finally, in Fig. \ref{nuCopt}, we plot the spatial throughput $\nu$ versus total cost $C$ for the cases with different $p$ and $K_N$, under their corresponding optimal IRS/BS density ratio $\zeta^*$.
First, it is observed that when the BS density is larger than a certain value, the optimal IRS/BS density ratio $\zeta^*$ is roughly proportional to the BS/IRS cost ratio $K_N$ and the network loading factor $p$.
Second, given the optimal ratio $\zeta^*$, the spatial throughput always increases with the total cost, where the speed of such increase is faster under a larger BS/IRS cost ratio $K_N$ or network loading factor $p$, due to similar reasons provided for Fig. \ref{nuRatio}.
The above results demonstrate that the IRS-aided hybrid (active/passive) wireless network can significantly enhance the network throughput as compared to the conventional network with active BSs only, when the densities of BSs and IRSs are optimally set based on their total cost constraint as well as other relevant network parameters.

\section{Conclusions}\label{SectionConclusion}
This paper investigates a new hybrid active/passive wireless network with large-scale deployment of BSs and IRSs, and proposes a new analytical framework based on stochastic geometry and probability theory to characterize its spatial throughput as well as other key performance metrics averaged over both random channel fading and BS/IRS locations.
Extensive numerical results are provided to validate our analysis and show the effectiveness of deploying IRSs to significantly enhance the signal power but with only marginally increased interference, thus greatly improving the network throughput as compared to the traditional wireless network with active BSs only, especially when the BS density and network loading factor are large.
Furthermore, it is shown that the new hybrid network with optimal IRS/BS density ratio can achieve a linear capacity growth with the network deployment cost, thus providing a fundamentally new approach to achieve sustainable capacity growth for future wireless networks.
Finally, incorporating multi-antenna BSs in the IRS-aided hybrid network can further enhance the network performance, which is worth investigating in future work. Nevertheless, this paper unveils that the new hybrid wireless network consisting of active BSs with single antenna only (thus much less costly than the conventional massive MIMO BSs) and passive IRSs can already achieve network capacity scaling cost-effectively, which thus provides an appealing alternative architecture for wireless networks, especially for their migration to higher frequency bands in the future.

\section*{Acknowledgement}
The authors would like to thank Dr. Qingqing Wu and Dr. Weidong Mei for their helps and the anonymous reviewers for their suggestions.

\appendices

\section{Mean and Variance of $h_{ir,m,n}^{(j)}$ for the IRS Random Scattering Case}\label{AppendixCSCG}
For the cascaded BS $m$-IRS $j$-UE 0 link in \eqref{hirmAll}, the channel $h_{\textrm{ir},m,n}^{(j)}$ of the reflected path via each element $n$ is given by \eqref{hirmn}, with the amplitude $A\triangleq |h_{\textrm{i},m,n}^{(j)}| |h_{\textrm{r},n}^{(j)}|$ and phase $\psi\triangleq \phi_n^{(j)}+\angle h_{\textrm{i},m,n}^{(j)}+\angle h_{\textrm{r},n}^{(j)}$.
For the case without passive beamforming,
the phase $\psi$ is uniformly random in $[0,2\pi)$ while the amplitude $A$ follows the double-Rayleigh distribution with mean and variance given by \eqref{hirmean} and \eqref{hirvariance}, respectively. Note that $A$ and $\psi$ are independent.
Denote $X\triangleq A\cos\psi$ and $Y\triangleq A\sin\psi$ as the in-phase and quadrature-phase components of $h_{\textrm{ir},m,n}^{(j)}$, respectively.
In the following, we derive the mean and variance of $X$, while those of $Y$ can be obtained similarly.

Due to the uniformly random $\psi$ in $[0,2\pi)$, the first two moments of $X$ are given by $\mathbb{E}\{X\}=\mathbb{E}\{A\cos\psi\}=0$ and $\mathbb{E}\{X^2\}=\mathbb{E}\{A^2\cos^2\psi\}=\mathbb{E}\{A^2\}\mathbb{E}\{\frac{1+\cos 2\psi}{2}\}=\frac{1}{2} \mathbb{E}\{A^2\}$, respectively. As a result, the mean of $X$ is 0 and its variance is given by
\begin{small}
\begin{align}
	&\var\{X\}=\mathbb{E}\{X^2\}-(\mathbb{E}\{X\})^2=\frac{1}{2} \mathbb{E}\{A^2\}=\frac{ \var\{A\}+(\mathbb{E}\{A\})^2}{2} \notag\\
	&=\frac{1}{2} \bigg( \big(1-\frac{\pi^2}{16}\big)g_{\textrm{i},m}^{(j)}g_{\textrm{r}}^{(j)}+\frac{\pi^2}{16}\cdot g_{\textrm{i},m}^{(j)}g_{\textrm{r}}^{(j)} \bigg)=\frac{1}{2} g_{\textrm{i},m}^{(j)}g_{\textrm{r}}^{(j)}.
\end{align}
\end{small}

\section{First and Second Moments of $h_1$, $h_2$, $|h_1|^2$ and $|h_2|^2$}\label{AppendixMoments}

For $h_1$, under given $l_0$ and $d_0$, we have
\begin{align}\label{h1}
&\mathbb{E}\{h_1\}|_{l_0,d_0}= \mathbb{E}\big\{ (|h_{\textrm{d},0}|+ |h_{\textrm{ir},0}^{(0)}|)e^{\bold i\angle h_{\textrm{d},0}} \big\}\big|_{l_0,d_0}\notag\\
&=\mathbb{E}\big\{|h_{\textrm{d},0}|+ |h_{\textrm{ir},0}^{(0)}|\big\}\big|_{l_0,d_0}\mathbb{E}\big\{e^{\bold i\angle h_{\textrm{d},0}}\big\}\big|_{l_0,d_0}=0,
\end{align}
due to $\mathbb{E}\big\{e^{\bold i\angle h_{\textrm{d},0}}\big\}\big|_{l_0,d_0}=0$.
For $h_2$, it is the sum of independent CSCG RVs and hence is CSCG distributed with zero mean, i.e., $\mathbb{E}\{h_2\}=0$. 
Based on similar derivations in \eqref{h1}, we have $\mathbb{E}\{h_1^2\}|_{l_0,d_0}=0$ and $\mathbb{E}\{h_2^2\}|_{l_0,d_0}=0$.

For $|h_1|^2$, its first moment conditioned on $l_0$ and $d_0$ is given by
\begin{small}
\begin{align}\label{h1square}
&\mathbb{E}\{|h_1|^2\}|_{l_0,d_0}= \mathbb{E}\{(|h_{\textrm{d},0}|+|h_{\textrm{ir},0}^{(0)}|)^2\}|_{l_0,d_0}\notag\\
&= \mathbb{E}\{|h_{\textrm{d},0}|^2+ 2 |h_{\textrm{d},0}| |h_{\textrm{ir},0}^{(0)}|+|h_{\textrm{ir},0}^{(0)}|^2\}|_{l_0,d_0}= \mathbb{E}\{|h_{\textrm{d},0}|^2\}|_{l_0,d_0}+ \notag\\
& 2 \mathbb{E}\{|h_{\textrm{d},0}|\}|_{l_0,d_0} \mathbb{E}\{|h_{\textrm{ir},0}^{(0)}|\}|_{l_0,d_0}+\mathbb{E}\{|h_{\textrm{ir},0}^{(0)}|^2\}|_{l_0,d_0}\notag\\
&=g_{\textrm{d},0}+2\cdot \sqrt{\frac{g_{\textrm{d},0}}{2}}\sqrt{\frac{\pi}{2}} \cdot \mathbb{E}_\varphi\big\{N\frac{\pi}{4}\sqrt{g_{\textrm{i},0}^{(0)} g_{\textrm{r}}^{(0)}}\big\} +\mathbb{E}_\varphi\big\{G_\textrm{bf} g_{\textrm{i},0}^{(0)} g_{\textrm{r}}^{(0)}\big\}\notag\\
&=g_{\textrm{d},0}+N\frac{\pi}{4}\sqrt{\pi  g_{\textrm{r}}^{(0)}g_{\textrm{d},0}}\mathbb{E}_\varphi\big\{\sqrt{g_{\textrm{i},0}^{(0)}}\big\} +G_\textrm{bf} g_{\textrm{r}}^{(0)}\mathbb{E}_\varphi\big\{g_{\textrm{i},0}^{(0)}\big\},
\end{align}
\end{small}%
where $g_{\textrm{i},0}^{(0)}$ is the average channel power gain of the BS 0-IRS 0 link with horizontal link distance $r_{0,0}=\sqrt{l_0^2+d_0^2-2l_0 d_0\cos\varphi}$ and $\varphi$ being the BS 0-UE 0-IRS $0$ angle projected on the ground plane.
The expectation of functions of $g_{\textrm{i},0}^{(0)}$ can thus be obtained by simple integrals over $\varphi$ in $[0,2\pi)$.
Similarly, we can obtain exact expressions (though not expressible in closed-form) for the moments of $|h_1|^2$, $|h_2|^2$ and hence the signal power $S$ conditioned on $l_0$ and $d_0$, by simple integrals over $\varphi$ and/or $d_j$.

Nevertheless, for simplicity, we apply the approximation of $g_{\textrm{i},0}^{(j)}\approx g_{\textrm{d},0}$, $j\in\mathcal{J}$ (discussed in the second paragraph of Section \ref{SectionSignal}) in the above expressions to obtain closed-form approximations for the moments of $S|_{l_0,d_0}$. 
Specifically, by applying $g_{\textrm{i},0}^{(j)}\approx g_{\textrm{d},0}=g_{\textrm{d}}(l_0)$ and $g_{\textrm{r}}^{(j)}=g_{\textrm{r}}(d_j)$ given by \eqref{gd} and \eqref{gr}, respectively, $\mathbb{E}\{|h_1|^2\}|_{l_0,d_0}$ in \eqref{h1square} is approximated by
\begin{equation}
\mathbb{E}\{|h_1|^2\}|_{l_0,d_0}\approx g_{\textrm{d}}(l_0)\bigg(1+N\frac{\pi}{4}\sqrt{\pi g_{\textrm{r}}(d_0)}+G_\textrm{bf} g_{\textrm{r}}(d_0)\bigg).
\end{equation}
Similarly, we have
\begin{small}
\begin{align}\label{h1_4}
&\mathbb{E}\{|h_1|^4\}|_{l_0,d_0}= \mathbb{E}\{(|h_{\textrm{d},0}|+|h_{\textrm{ir},0}^{(0)}|)^4\}|_{l_0,d_0}= \mathbb{E}\big\{|h_{\textrm{d},0}|^4+\notag\\
&4|h_{\textrm{d},0}|^3|h_{\textrm{ir},0}^{(0)}|+6|h_{\textrm{d},0}|^2|h_{\textrm{ir},0}^{(0)}|^2+4|h_{\textrm{d},0}||h_{\textrm{ir},0}^{(0)}|^3+|h_{\textrm{ir},0}^{(0)}|^4\big\}|_{l_0,d_0}\notag\\
&\approx [g_\textrm{d}(l_0)]^2\bigg[ 2  +  \frac{3}{4}\pi^{\frac{3}{2}}N \sqrt{g_\textrm{r}(d_0)}  +  6 G_\textrm{bf} g_\textrm{r}(d_0)  +\notag\\
&  2\sqrt{\pi}\bigg( \frac{\pi^3 N^3}{64}  +\frac{3 \pi N^2 (1-\frac{\pi^2}{16})}{4} \bigg)[g_\textrm{r}(d_0)]^{\frac{3}{2}}+  \notag\\
& \bigg(\frac{\pi^4 N^4}{256} + \frac{3\pi^2 N^3 (1-\frac{\pi^2}{16}) }{8}  + 3N^2 \big(1-\frac{\pi^2}{16}\big)^2  \bigg) [g_\textrm{r}(d_0)]^2 \bigg].
\end{align}
\end{small}


Finally, under given BS/IRS locations, $h_2$ is CSCG distributed with zero mean and covariance $\sum_{j\in\mathcal{J}\setminus\{0\}}N g_{\textrm{i},0}^{(j)}g_{\textrm{r}}^{(j)}$, and thus $|h_2|^2$ follows the exponential distribution with mean $\sum_{j\in\mathcal{J}\setminus\{0\}}N g_{\textrm{i},0}^{(j)}g_{\textrm{r}}^{(j)}$.
Therefore, the two moments of $|h_2|^2$ conditioned on $l_0$ and $d_0$ are respectively approximated by
	\begin{align}\label{h2square}
	&\mathbb{E}\{|h_2|^2\}|_{l_0,d_0}=\mathbb{E}\bigg\{\sum_{j\in\mathcal{J}\setminus\{0\}}N g_{\textrm{i},0}^{(j)}g_{\textrm{r}}^{(j)}\bigg\}\bigg|_{l_0,d_0}\notag\\
	&\approx \mathbb{E}\bigg\{ N g_{\textrm{d}}(l_0) \sum_{j\in\mathcal{J}\setminus\{0\}} g_{\textrm{r}}(d_j)\bigg\}\bigg|_{l_0,d_0}=N g_{\textrm{d}}(l_0) E_\textrm{I1}(d_0),
	\end{align}
	\begin{align}\label{h2_4}
	&\mathbb{E}\{|h_2|^4\}|_{l_0,d_0}=\mathbb{E}\bigg\{2\bigg(\sum_{j\in\mathcal{J}\setminus\{0\}}N g_{\textrm{i},0}^{(j)}g_{\textrm{r}}^{(j)}\bigg)^2\bigg\}\bigg|_{l_0,d_0} \notag\\
	&\approx 2 N^2 [g_{\textrm{d}}(l_0)]^2 \mathbb{E}\bigg\{\bigg( \sum_{j\in\mathcal{J}\setminus\{0\}} g_{\textrm{r}}(d_j)\bigg)^2\bigg\}\bigg|_{d_0}\notag\\
	&=2 N^2 [g_{\textrm{d}}(l_0)]^2E_\textrm{I3}(d_0),
	\end{align}%
where $E_\textrm{I1}(d_0)$ and $E_\textrm{I3}(d_0)$ are given in \eqref{EI1} and \eqref{EI3}, respectively.

\section{Derivation of Interference Power Laplace Transform}\label{AppendixLaplace}

The Laplace transform of the interference power conditioned on $l_0$ and $d_0$ is given by
\begin{align}
&\mathcal{L}_{I|_{l_0,d_0}}(s)\triangleq \mathbb{E}\{e^{-sI}\}|_{l_0,d_0}\notag\\
&\approx\mathbb{E}\bigg\{\exp\bigg(-s\bar\eta\sum_{m\in \Lambda_\textrm{B}'\setminus\{0\}} g_{\textrm{d}}(l_m) \xi_m\bigg)\bigg\}\bigg|_{l_0,d_0}\notag\\
&\stackrel{(c)}{=}\mathbb{E}_{\Lambda_\textrm{B}'}\bigg\{\prod_{m\in \Lambda_\textrm{B}'\setminus\{0\}}\mathbb{E}_{\xi}\big\{\exp\big(-s\bar\eta g_{\textrm{d}}(l_m) \xi\big)\big\}\bigg\}\bigg|_{l_0,d_0}\notag\\
&\stackrel{(d)}{=}\exp\bigg(-2\pi\lambda_\textrm{B}'\int_{l_0}^{\infty}\bigg[1-\mathbb{E}_{\xi}\big\{\exp\big(-s\bar\eta g_{\textrm{d}}(l) \xi\big)\big\}\bigg]l\diff l\bigg)\notag\\
&\stackrel{(e)}{=}\exp\bigg(-2\pi\lambda_\textrm{B}'\int_{l_0}^{\infty}\bigg[1-\frac{1}{1+s\bar\eta g_{\textrm{d}}(l)}\bigg]l\diff l\bigg)\notag\\
&=\exp\big(-2\pi\lambda_\textrm{B}' U( s\bar\eta) \big),
\end{align}
where $(c)$ follows from i.i.d. $\xi_m\stackrel{\textrm{dist.}}{=}\xi\sim \textrm{Exp}(1), \forall m$ and independent $\Lambda_\textrm{B}'$; $(d)$ is based on the probability generating functional of HPPP \cite{AndrewsCellular}; $(e)$ is due to the fact that $\mathbb{E}_{\xi}\{e^{-s\xi}\}\triangleq \frac{1}{1+s}$ for $\xi\sim \textrm{Exp}(1)$; and $U(\cdot)$ is defined in \eqref{Ux}.

\section{Proof of Lemma \ref{lem_Pno}}\label{AppendixPno}
For a Gamma-distributed RV $S\sim \Gamma[k_S,\theta_S]$, its complementary cdf (ccdf) is given by $\mathbb{P}\{S>x\}=\Gamma\big(k_S,\frac{x}{\theta_S}\big)/\Gamma(k_S)$. As a result, for an independent RV $X$, we have
\begin{equation}\label{PSX1}
	\mathbb{P}\{S>X\}=\mathbb{E}_{X}\bigg\{\frac{\Gamma\big(k_S,\frac{X}{\theta_S}\big)}{\Gamma(k_S)}\bigg\}.
\end{equation}
Let $Y=X/\theta_S$. Using the fact that $\Gamma\big(k_S, y\big)/\Gamma(k_S)=\sum_{i=0}^{k_S-1} y^i e^{-y}/i!$ for integer $k_S$ \cite{NISTfunctions}, we have
\begin{align}\label{PSX2}
\mathbb{P}\{S>X\}&=\mathbb{E}_{Y}\bigg\{\sum_{i=0}^{k_S-1} Y^i e^{-Y}/i!\bigg\}\notag\\
&=\sum_{i=0}^{k_S-1} \frac{(-1)^i}{i!}\mathbb{E}_{Y}\big\{(-1)^i Y^i e^{-Y}\big\}.
\end{align}
Based on the fact that $\frac{\partial^i}{\partial s^i}\big[e^{-sY}\big]=(-Y)^i e^{-sY}$ as well as the interchangeable operators of expectation and differentiation, we thus have
\begin{align}\label{PSX3}
\mathbb{P}\{S>X\}&=\sum_{i=0}^{k_S-1} \frac{(-1)^i}{i!}\mathbb{E}_{Y}\bigg\{\frac{\partial^i}{\partial s^i}\big[e^{-sY}\big]_{s=1}\bigg\}\notag\\
&=\sum_{i=0}^{k_S-1} \frac{(-1)^i}{i!}\frac{\partial^i}{\partial s^i}\bigg[\mathbb{E}_{Y}\big\{e^{-sY}\big\}\bigg]_{s=1}.
\end{align}
Therefore, Lemma \ref{lem_Pno} follows.

\section{Derivatives of the Composite Function $\exp\big(V(s)\big)$}\label{AppendixVs}

The high-order derivatives of the composite function $\exp\big(V(s)\big)$ can be evaluated efficiently using the Fa\`a di Bruno's formula\cite{AndrewsGeneralFading}. Since the outer function is an exponential function, we have 
\begin{equation}
\frac{\partial^i}{\partial s^i}\big[\exp\big(V(s)\big)\big]=\exp\big(V(s)\big) B_i\bigg(\frac{\partial^1 V(s)}{\partial s^1},\cdots,\frac{\partial^i V(s)}{\partial s^i}\bigg),
\end{equation}
where $B_i(x_1,\cdots,x_i)$ is the $i$-th complete Bell polynomial with fixed and known coefficients. It remains to compute the derivatives of the inner function $V(s)$ up to order $k_S-1$, which resorts to the derivatives of $U(x)$ in \eqref{Ux}.
Define $\tilde{x}\triangleq \frac{\bar \gamma \bar\eta}{\theta_S}$.
The 1st-order derivative of $V(s)$ is given by
\begin{equation}
\frac{\partial^1 V(s)}{\partial s^1}=-\frac{\bar \gamma W }{\theta_S} -2\pi\lambda_\textrm{B}' \tilde{x} \frac{\partial^1 U(x)}{\partial x^1}\bigg|_{x=s\tilde{x}}.
\end{equation}
For $i>1$, we have
\begin{equation}
\frac{\partial^i V(s)}{\partial s^i}=-2\pi\lambda_\textrm{B}' \tilde{x}^i \frac{\partial^i U(x)}{\partial x^i}\bigg|_{x=s\tilde{x}}.
\end{equation}
Finally, the high-order derivatives of $U(x)$ are given by
\begin{equation}
\frac{\partial^i U(x)}{\partial x^i}=  (\delta)_i b_1 x^{\delta-i} - b_2 \bigg(\frac{(\delta)_i H(x)}{x^i} +L_i(x)   \bigg), i\geq 1,
\end{equation}
where $\delta\triangleq \frac{2}{\alpha}$, $b_1\triangleq\frac{\pi \beta^{\frac{2}{\alpha}}}{\alpha \sin(\frac{2\pi}{\alpha})}$,  $b_2\triangleq\frac{l_0^2+H_\textrm{B}^2}{2}$, $b_3\triangleq g_\textrm{d}(l_0)$,
$H(x)\triangleq {}_{2}F_{1}\big(1,\delta,1+\delta,-1/(b_3 x)\big)$, $(\delta)_i\triangleq \delta(\delta-1)\cdots(\delta-i+1)$ is the falling factorial, and $L_i(x)$ are polynomial fractions that can be readily obtained via symbolic tools like Mathematica. For example, we have
\begin{equation}
L_1(x)\triangleq -\frac{\delta b_3 }{1+b_3 x},
\end{equation}
\begin{equation}
L_2(x)\triangleq -\frac{ \delta b_3  \big(\delta+b_3(\delta-1)x\big)}{x(1+b_3 x)^2},
\end{equation}
\begin{equation}\small
L_3(x)\triangleq -\frac{\delta b_3  \big(2b_3^2 x^2+\delta^2(1+b_3 x)^2-\delta(1+b_3 x)(2+3b_3 x)\big)}{x^2(1+b_3 x)^3}.
\end{equation}

\bibliography{IEEEabrv,BibDIRP}

\begin{thebibliography}{10}
\providecommand{\url}[1]{#1}
\csname url@samestyle\endcsname
\providecommand{\newblock}{\relax}
\providecommand{\bibinfo}[2]{#2}
\providecommand{\BIBentrySTDinterwordspacing}{\spaceskip=0pt\relax}
\providecommand{\BIBentryALTinterwordstretchfactor}{4}
\providecommand{\BIBentryALTinterwordspacing}{\spaceskip=\fontdimen2\font plus
\BIBentryALTinterwordstretchfactor\fontdimen3\font minus
  \fontdimen4\font\relax}
\providecommand{\BIBforeignlanguage}[2]{{%
\expandafter\ifx\csname l@#1\endcsname\relax
\typeout{** WARNING: IEEEtran.bst: No hyphenation pattern has been}%
\typeout{** loaded for the language `#1'. Using the pattern for}%
\typeout{** the default language instead.}%
\else
\language=\csname l@#1\endcsname
\fi
#2}}
\providecommand{\BIBdecl}{\relax}
\BIBdecl

\bibitem{SmallCell5G}
J.~G. Andrews \emph{et~al.}, ``What will {5G} be?'' \emph{IEEE J. Sel. Areas
  Commun.}, vol.~32, no.~6, pp. 1065--1082, June 2014.

\bibitem{QQirsMag}
Q.~Wu and R.~Zhang, ``Towards smart and reconfigurable environment:
  {Intelligent} reflecting surface aided wireless network,'' \emph{IEEE Commun.
  Mag.}, vol.~58, no.~1, pp. 106--112, Jan. 2020.

\bibitem{IRSbasar}
E.~{Basar}, M.~{Di Renzo}, J.~{De Rosny}, M.~{Debbah}, M.~{Alouini}, and
  R.~{Zhang}, ``Wireless communications through reconfigurable intelligent
  surfaces,'' \emph{IEEE Access}, vol.~7, pp. 116\,753--116\,773, Sept. 2019.

\bibitem{IRSmarcoCome}
M.~Di~Renzo, M.~Debbah, D.-T. Phan-Huy, A.~Zappone, M.-S. Alouini, C.~Yuen,
  V.~Sciancalepore, G.~C. Alexandropoulos, J.~Hoydis, H.~Gacanin, J.~De~Rosny,
  A.~Bounceur, G.~Lerosey, and M.~Fink, ``Smart radio environments empowered by
  reconfigurable {AI} meta-surfaces: {An} idea whose time has come,''
  \emph{EURASIP J. Wireless Commun. and Netw.}, vol. 2019, no.~1, pp. 1--20,
  2019.

\bibitem{IRSholographic}
C.~Huang, S.~Hu, G.~C. Alexandropoulos, A.~Zappone, C.~Yuen, R.~Zhang,
  M.~Di~Renzo, and M.~Debbah, ``Holographic {MIMO} surfaces for {6G} wireless
  networks: {Opportunities}, challenges, and trends,'' \emph{IEEE Wireless
  Commun.}, vol.~27, no.~5, pp. 118--125, 2020.

\bibitem{QQtwc}
Q.~{Wu} and R.~{Zhang}, ``Intelligent reflecting surface enhanced wireless
  network via joint active and passive beamforming,'' \emph{IEEE Trans.
  Wireless Commun.}, vol.~18, no.~11, pp. 5394--5409, Nov. 2019.

\bibitem{IRSjinShi}
Y.~{Han}, W.~{Tang}, S.~{Jin}, C.~{Wen}, and X.~{Ma}, ``Large intelligent
  surface-assisted wireless communication exploiting statistical {CSI},''
  \emph{IEEE Trans. Veh. Technol.}, vol.~68, no.~8, pp. 8238--8242, Aug. 2019.

\bibitem{IRSschoberICCC}
X.~{Yu}, D.~{Xu}, and R.~{Schober}, ``{MISO} wireless communication systems via
  intelligent reflecting surfaces,'' in \emph{Proc. IEEE/CIC Int. Conf. Commun.
  China (ICCC)}, Aug. 2019, pp. 735--740.

\bibitem{IRSqqDiscrete}
Q.~Wu and R.~Zhang, ``Beamforming optimization for wireless network aided by
  intelligent reflecting surface with discrete phase shifts,'' \emph{IEEE
  Trans. Commun.}, vol.~68, no.~3, pp. 1838--1851, Mar. 2020.

\bibitem{IRSshuowen}
S.~{Zhang} and R.~{Zhang}, ``Capacity characterization for intelligent
  reflecting surface aided {MIMO} communication,'' \emph{IEEE J. Sel. Areas
  Commun.}, vol.~38, no.~8, pp. 1823--1838, 2020.

\bibitem{IRScuiYing}
\BIBentryALTinterwordspacing
Z.~Zhang, Y.~Cui, F.~Yang, and L.~Ding, ``Analysis and optimization of outage
  probability in multi-intelligent reflecting surface-assisted systems,'' 2019.
  [Online]. Available: \url{https://arxiv.org/abs/1909.02193}
\BIBentrySTDinterwordspacing

\bibitem{IRSchauYuen}
C.~{Huang}, A.~{Zappone}, G.~C. {Alexandropoulos}, M.~{Debbah}, and C.~{Yuen},
  ``Reconfigurable intelligent surfaces for energy efficiency in wireless
  communication,'' \emph{IEEE Trans. Wireless Commun.}, vol.~18, no.~8, pp.
  4157--4170, Aug. 2019.

\bibitem{IRSyifeiOFDM}
Y.~{Yang}, B.~{Zheng}, S.~{Zhang}, and R.~{Zhang}, ``Intelligent reflecting
  surface meets {OFDM: Protocol} design and rate maximization,'' \emph{IEEE
  Trans. Commun.}, vol.~68, no.~7, pp. 4522--4535, 2020.

\bibitem{IRSzhengBeixiong}
B.~Zheng and R.~Zhang, ``Intelligent reflecting surface-enhanced {OFDM:
  Channel} estimation and reflection optimization,'' \emph{IEEE Wireless
  Commun. Lett.}, vol.~9, no.~4, pp. 518--522, 2020.

\bibitem{IRSyangGangNOMA}
G.~{Yang}, X.~{Xu}, and Y.~{Liang}, ``Intelligent reflecting surface assisted
  non-orthogonal multiple access,'' in \emph{Proc. IEEE Wireless Commun. Netw.
  Conf. (WCNC)}, 2020, pp. 1--6.

\bibitem{IRSdingZhiguoNOMA}
Z.~{Ding} and H.~{Vincent Poor}, ``A simple design of {IRS-NOMA}
  transmission,'' \emph{IEEE Commun. Lett.}, vol.~24, no.~5, pp. 1119--1123,
  2020.

\bibitem{IRSbeixiongNOMA}
B.~{Zheng}, Q.~{Wu}, and R.~{Zhang}, ``Intelligent reflecting surface-assisted
  multiple access with user pairing: {NOMA or OMA}?'' \emph{IEEE Commun.
  Lett.}, vol.~24, no.~4, pp. 753--757, 2020.

\bibitem{IRSsecureGuangchi}
M.~{Cui}, G.~{Zhang}, and R.~{Zhang}, ``Secure wireless communication via
  intelligent reflecting surface,'' \emph{IEEE Wireless Commun. Lett.}, vol.~8,
  no.~5, pp. 1410--1414, Oct. 2019.

\bibitem{IRSsecureYCliang}
J.~{Chen}, Y.~{Liang}, Y.~{Pei}, and H.~{Guo}, ``Intelligent reflecting
  surface: {A} programmable wireless environment for physical layer security,''
  \emph{IEEE Access}, vol.~7, pp. 82\,599--82\,612, 2019.

\bibitem{IRSsecureSchober}
D.~Xu, X.~Yu, Y.~Sun, D.~W.~K. Ng, and R.~Schober, ``Resource allocation for
  secure {IRS}-assisted multiuser {MISO} systems,'' in \emph{IEEE Proc.
  GLOBECOM Workshops}, 2019, pp. 1--6.

\bibitem{IRSsecureXinrong}
X.~Guan, Q.~Wu, and R.~Zhang, ``Intelligent reflecting surface assisted secrecy
  communication: {Is} artificial noise helpful or not?'' \emph{IEEE Wireless
  Commun. Lett.}, vol.~9, no.~6, pp. 778--782, 2020.

\bibitem{QQirsSWIPTwcl}
Q.~{Wu} and R.~{Zhang}, ``Weighted sum power maximization for intelligent
  reflecting surface aided {SWIPT},'' \emph{IEEE Wireless Commun. Lett.},
  vol.~9, no.~5, pp. 586--590, 2020.

\bibitem{IRSnalanSWIPT}
C.~{Pan}, H.~{Ren}, K.~{Wang}, M.~{Elkashlan}, A.~{Nallanathan}, J.~{Wang}, and
  L.~{Hanzo}, ``Intelligent reflecting surface aided {MIMO} broadcasting for
  simultaneous wireless information and power transfer,'' \emph{IEEE J. Sel.
  Areas Commun.}, vol.~38, no.~8, pp. 1719--1734, 2020.

\bibitem{QQirsSWIPTqos}
Q.~{Wu} and R.~{Zhang}, ``Joint active and passive beamforming optimization for
  intelligent reflecting surface assisted {SWIPT Under QoS} constraints,''
  \emph{IEEE J. Sel. Areas Commun.}, vol.~38, no.~8, pp. 1735--1748, 2020.

\bibitem{HanzoMIMOstochasticGeometry}
\BIBentryALTinterwordspacing
T.~Hou, Y.~Liu, Z.~Song, X.~Sun, Y.~Chen, and L.~Hanzo, ``{MIMO} assisted
  networks relying on large intelligent surfaces: {A} stochastic geometry
  model,'' 2019. [Online]. Available: \url{https://arxiv.org/abs/1910.00959}
\BIBentrySTDinterwordspacing

\bibitem{IRSlyuSingleCell}
J.~Lyu and R.~Zhang, ``Spatial throughput characterization for intelligent
  reflecting surface aided multiuser system,'' \emph{IEEE Wireless Commun.
  Lett.}, vol.~9, no.~6, pp. 834--838, 2020.

\bibitem{IRSmulticellNallan}
C.~{Pan}, H.~{Ren}, K.~{Wang}, W.~{Xu}, M.~{Elkashlan}, A.~{Nallanathan}, and
  L.~{Hanzo}, ``Multicell {MIMO} communications relying on intelligent
  reflecting surfaces,'' \emph{IEEE Trans. Wireless Commun.}, vol.~19, no.~8,
  pp. 5218--5233, 2020.

\bibitem{IRSmulticellXujie}
H.~{Xie}, J.~{Xu}, and Y.~{Liu}, ``Max-min fairness in {IRS}-aided multi-cell
  {MISO} systems with joint transmit and reflective beamforming,'' in
  \emph{Proc. IEEE Int. Conf. Commun. (ICC)}, 2020, pp. 1--6.

\bibitem{IRSmulticellCuiYing}
Y.~Jia, C.~Ye, and Y.~Cui, ``Analysis and optimization of an intelligent
  reflecting surface-assisted system with interference,'' in \emph{Proc. IEEE
  Int. Conf. Commun. (ICC)}, 2020, pp. 1--6.

\bibitem{IRSdistributedMultiPairsShiYuanming}
J.~{He}, K.~{Yu}, and Y.~{Shi}, ``Coordinated passive beamforming for
  distributed intelligent reflecting surfaces network,'' in \emph{Proc. IEEE
  Veh. Technol. Conf.}, 2020, pp. 1--5.

\bibitem{MarcoReflection}
M.~Di~Renzo and J.~Song, ``Reflection probability in wireless networks with
  metasurface-coated environmental objects: {An} approach based on random
  spatial processes,'' \emph{EURASIP J. Wireless Commun. Netw.}, vol. 2019,
  no.~1, p.~99, 2019.

\bibitem{IRSalouiniBlockage}
M.~A. Kishk and M.-S. Alouini, ``Exploiting randomly located blockages for
  large-scale deployment of intelligent surfaces,'' \emph{IEEE J. Sel. Areas
  Commun.}, vol.~39, no.~4, pp. 1043--1056, 2021.

\bibitem{AndrewsCellular}
J.~G. Andrews, F.~Baccelli, and R.~K. Ganti, ``A tractable approach to coverage
  and rate in cellular networks,'' \emph{IEEE Trans. Commun.}, vol.~59, no.~11,
  pp. 3122--3134, Nov. 2011.

\bibitem{SGexperiment}
W.~Lu and M.~Di~Renzo, ``Stochastic geometry modeling of cellular networks:
  {Analysis}, simulation and experimental validation,'' in \emph{Proc. ACM Int.
  Conf. MSWiM}, 2015, p. 179–188.

\bibitem{AndrewsPrimer}
\BIBentryALTinterwordspacing
J.~G. Andrews, A.~K. Gupta, and H.~S. Dhillon, ``A primer on cellular network
  analysis using stochastic geometry,'' 2016. [Online]. Available:
  \url{https://arxiv.org/abs/1604.03183}
\BIBentrySTDinterwordspacing

\bibitem{forbes2011statistical}
C.~Forbes, M.~Evans, N.~Hastings, and B.~Peacock, \emph{Statistical
  distributions}.\hskip 1em plus 0.5em minus 0.4em\relax John Wiley \& Sons,
  2011.

\bibitem{DoubleRayleigh}
J.~{Salo}, H.~M. {El-Sallabi}, and P.~{Vainikainen}, ``The distribution of the
  product of independent {Rayleigh} random variables,'' \emph{IEEE Trans.
  Antennas Propag.}, vol.~54, no.~2, pp. 639--643, 2006.

\bibitem{NISTfunctions}
F.~W. Olver, D.~W. Lozier, R.~F. Boisvert, and C.~W. Clark, \emph{NIST handbook
  of mathematical functions}.\hskip 1em plus 0.5em minus 0.4em\relax Cambridge
  university press, 2010.

\bibitem{RobertHeathGamma}
R.~W. {Heath}, M.~{Kountouris}, and T.~{Bai}, ``Modeling heterogeneous network
  interference using {Poisson} point processes,'' \emph{IEEE Trans. Signal
  Process.}, vol.~61, no.~16, pp. 4114--4126, Aug. 2013.

\bibitem{AndrewsGeneralFading}
R.~{Tanbourgi}, H.~S. {Dhillon}, J.~G. {Andrews}, and F.~K. {Jondral},
  ``Dual-branch {MRC} receivers under spatial interference correlation and
  {Nakagami} fading,'' \emph{IEEE Trans. Commun.}, vol.~62, no.~6, pp.
  1830--1844, June 2014.

\end{thebibliography}

\vspace{-0.3in}
\begin{IEEEbiography}[{\includegraphics[width=1in,height=1.25in,clip,keepaspectratio]{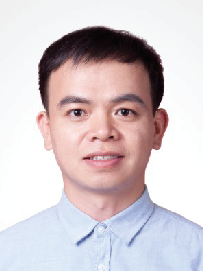}}]{Jiangbin Lyu}
	(S'12, M'16) received his B. Eng. degree (Hons.) in control science and engineering (under the Chu Kochen Honors Program) from Zhejiang University, Hangzhou, China, in 2011, and the Ph.D. degree from NUS Graduate School for Integrative Sciences and Engineering (NGS) (under the NGS scholarship), National University of Singapore (NUS), Singapore, in 2015. 
	
    He was a Post-Doctoral Research Fellow with the Department of Electrical and Computer Engineering, NUS, from 2015 to 2017. He is currently an assistant professor in the School of Informatics, Xiamen University, China, with research interests in unmanned aerial vehicle communications, intelligent reflecting surface, cross-layer network optimization, etc. 
        
    Dr. Lyu was a recipient of the IEEE Communications Society Heinrich Hertz Prize Paper Award in 2020, and also the Best Paper Award at Singapore-Japan International Workshop on Smart Wireless Communications in 2014. He served as the Invited Track Co-Chair at the 2021 IEEE/CIC ICCC conference, a TPC member for IEEE GLOBECOM and ICC, and a reviewer for various IEEE journals.	
\end{IEEEbiography}

\begin{IEEEbiography}[{\includegraphics[width=1in,height=1.25in,clip,keepaspectratio]{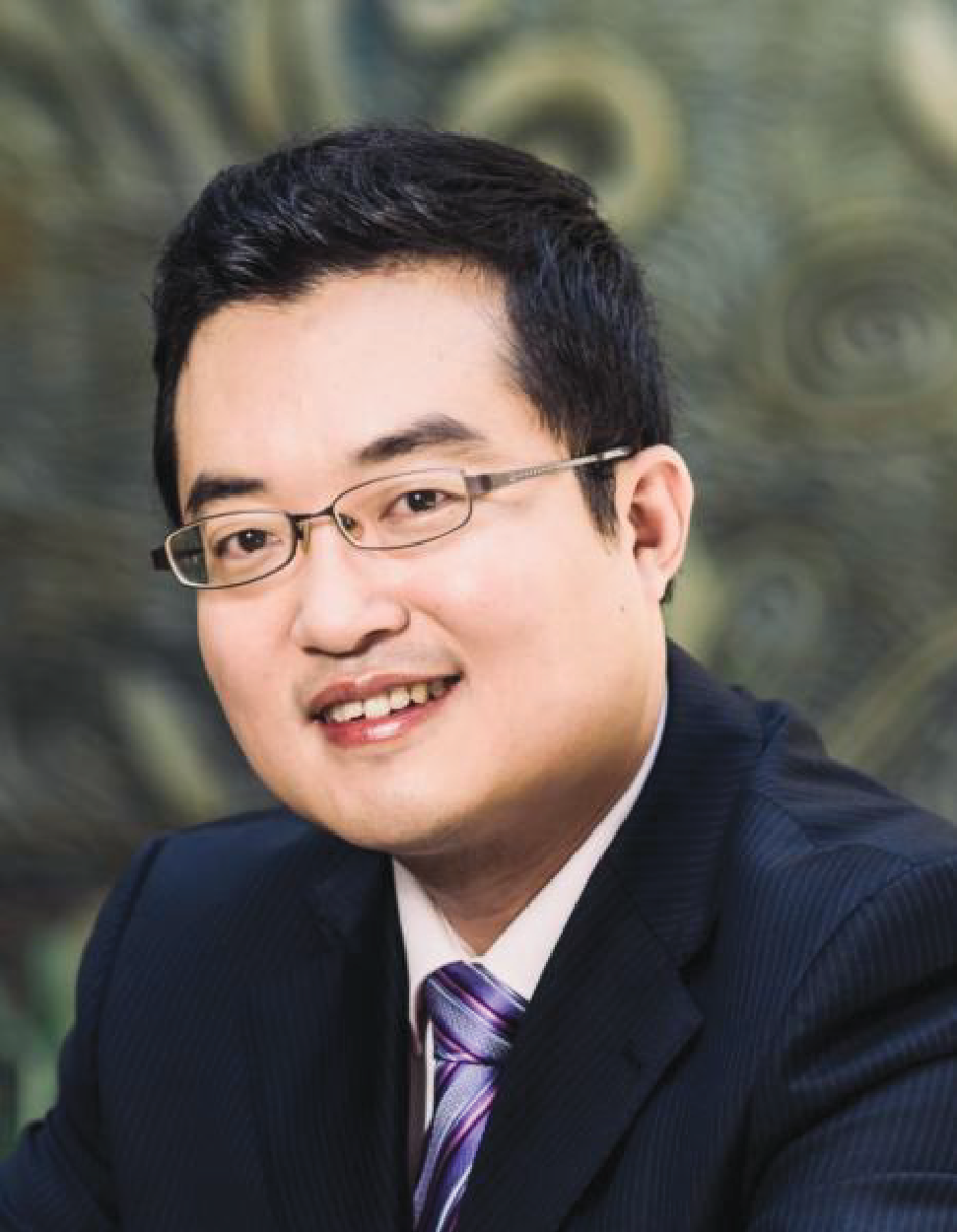}}] {Rui Zhang} (S'00-M'07-SM'15-F'17) received the B.Eng. (first-class Hons.) and M.Eng. degrees from the National University of Singapore, Singapore, and the Ph.D. degree from the Stanford University, Stanford, CA, USA, all in electrical engineering.
	
	From 2007 to 2010, he worked at the Institute for Infocomm Research, ASTAR, Singapore. Since 2010, he has been working with the National University of Singapore, where he is now a Provost’s Chair Professor in the Department of Electrical and Computer Engineering. He has published over 250 journal papers and over 190 conference papers. He has been listed as a Highly Cited Researcher by Thomson Reuters/Clarivate Analytics since 2015. His current research interests include UAV/satellite communications, wireless power transfer, reconfigurable MIMO, and optimization methods.     
	
	He was the recipient of the 6th IEEE Communications Society Asia-Pacific Region Best Young Researcher Award in 2011, the Young Researcher Award of National University of Singapore in 2015, the Wireless Communications Technical Committee Recognition Award in 2020, and the IEEE Signal Processing and Computing for Communications (SPCC) Technical Recognition Award in 2020. He was the co-recipient of the IEEE Marconi Prize Paper Award in Wireless Communications in 2015 and 2020, the IEEE Communications Society Asia-Pacific Region Best Paper Award in 2016, the IEEE Signal Processing Society Best Paper Award in 2016, the IEEE Communications Society Heinrich Hertz Prize Paper Award in 2017 and 2020, the IEEE Signal Processing Society Donald G. Fink Overview Paper Award in 2017, and the IEEE Communications Society Stephen O. Rice Prize in 2021. His co-authored paper received the IEEE Signal Processing Society Young Author Best Paper Award in 2017, and the IEEE Communications Society Young Author Best Paper Award in 2021. He served for over 30 international conferences as the TPC co-chair or an organizing committee member, and as the guest editor for 3 special issues in the IEEE JOURNAL OF SELECTED TOPICS IN SIGNAL PROCESSING and the IEEE JOURNAL ON SELECTED AREAS IN COMMUNICATIONS. He was an elected member of the IEEE Signal Processing Society SPCOM Technical Committee from 2012 to 2017 and SAM Technical Committee from 2013 to 2015, and served as the Vice Chair of the IEEE Communications Society Asia-Pacific Board Technical Affairs Committee from 2014 to 2015. He served as an Editor for the IEEE TRANSACTIONS ON WIRELESS COMMUNICATIONS from 2012 to 2016, the IEEE JOURNAL ON SELECTED AREAS IN COMMUNICATIONS: Green Communications and Networking Series from 2015 to 2016, the IEEE TRANSACTIONS ON SIGNAL PROCESSING from 2013 to 2017, and the IEEE TRANSACTIONS ON GREEN COMMUNICATIONS AND NETWORKING from 2016 to 2020. He is now an Editor for the IEEE TRANSACTIONS ON COMMUNICATIONS. He serves as a member of the Steering Committee of the IEEE Wireless Communications Letters. He served as a Distinguished Lecturer of IEEE Signal Processing Society and IEEE Communications Society from 2019 to 2020. \end{IEEEbiography}

\end{document}